\documentclass[article,twocolumn,prd,superscriptaddress,preprintnumbers,nofootinbib]{revtex4-2}
\usepackage[utf8]{inputenc}
\usepackage{amsmath,amsthm,amssymb,mathtools,physics,cancel,mhchem,tikz}
\usepackage{graphicx,flexisym,float,mwe}
\usepackage{dcolumn}
\usepackage[hidelinks]{hyperref}
\hypersetup{colorlinks=true,
    linkcolor=blue,
    filecolor=blue,
    urlcolor=blue,
    citecolor=blue,
    }

\usepackage{caption}
\usepackage{subcaption}
\usepackage{adjustbox}
\usepackage{gensymb}
\usepackage{breqn}
\usepackage{braket}
\usepackage{enumitem}
\usepackage{gensymb}

\begin{document}

\hfill \preprint{MI-HET-776}
\hfill \preprint{FERMILAB-PUB-22-485-ND}

\title{Inelastic nuclear scattering from neutrinos and dark matter}

\author{Bhaskar Dutta}
\email{dutta@physics.tamu.edu}

\author{Wei-Chih Huang}
\email{s104021230@tamu.edu}
\affiliation{Mitchell Institute for Fundamental Physics and Astronomy$,$ Department~ of ~Physics ~ and~ Astronomy$,$\\ Texas A$\&$M University$,$~College~ Station$,$ ~Texas ~77843$,$~ USA}

\author{Jayden L.~Newstead}
\email{jnewstead@unimelb.edu.au}
\affiliation{ARC Centre of Excellence for Dark Matter Particle Physics$,$ \\~School of Physics$,$~ The~ University~ of~ Melbourne$,$~ Victoria~ 3010$,$~ Australia}

\author{Vishvas Pandey}
\email{vpandey@fnal.gov}
\affiliation{Fermi National Accelerator Laboratory$,$ Batavia$,$ Illinois 60510$,$ USA}
\affiliation{Department of Physics$,$ University of Florida$,$ Gainesville$,$ FL 32611$,$ USA}

\begin{abstract}
Neutrinos with energy of order 10~MeV, such as from pion decay-at-rest sources, are an invaluable tool for studying low-energy neutrino interactions with nuclei -- previously enabling the first measurement of coherent elastic neutrino-nucleus scattering. Beyond elastic scattering, neutrinos and dark matter in this energy range also excite nuclei to its low-lying nuclear states, providing an additional physics channel. Here, we consider neutral-current inelastic neutrino-nucleus and dark matter(DM)-nucleus scattering off $^{40}$Ar, $^{133}$Cs, and $^{127}$I nuclei that are relevant to a number of low-threshold neutrino experiments at pion decay-at-rest facilities. We carry out large scale nuclear shell model calculations of the inelastic cross sections considering the full set of electroweak multipole operators. Our results demonstrate that Gamow-Teller transitions provide the dominant contribution to the cross section and that the long-wavelength limit provides a reasonable approximation to the total cross section for neutrino sources. We show that future experiments will be sensitive to this channel and thus these results provide additional neutrino and DM scattering channels to explore at pion decay-at-rest facilities.
\end{abstract}

\maketitle

\section{Introduction}
The large flux of neutrinos produced as a by-product at spallation neutron sources has enabled new tests of the Standard Model of particle physics. Most notably, the first measurement of coherent elastic neutrino nucleus scattering (CE$\nu$NS) by the COHERENT collaboration at the Spallation Neutron Source (SNS) in 2017~\cite{Akimov:2017ade}. Several other CE$\nu$NS experiments have been completed or are in operation, such as Coherent CAPTAIN-Mills (CCM) at Los Alamos National Laboratory (LANL) \cite{CCM}. Additional measurements have improved in precision and thus far they are in agreement with the Standard Model (SM) predictions~\cite{Akimov:2021dab}. Further improvements in precision will allow these experiments to probe physics beyond the SM, e.g. non-standard neutrino interactions~\cite{Ohlsson:2012kf,Miranda:2015dra,Dent:2017mpr,Liao:2017uzy, Denton:2018xmq,Billard:2018jnl, Altmannshofer:2018xyo, Dutta:2019eml,Canas:2019fjw, Khan:2019cvi,Coloma:2020nhf, Denton:2020hop, Dutta:2020che} and dark matter~\cite{deNiverville:2011it,deNiverville:2015mwa,Ge:2017mcq,Dutta:PRL2020,COHERENT:2021pvd,COHERENT:2019kwz,Aguilar-Arevalo:2021sbh,CCM:2021leg}. 

Other SM signals that could be observable at these experiments include inelastic neutrino-nucleus scattering, where the nucleus is left in an excited state. Neutrinos with energies of tens of MeV can excite many states in the various target nuclei used for these experiments. While the cross section for inelastic scattering is much smaller than the coherently enhanced elastic scattering, these processes could provide unique tests of new physics and/or important backgrounds to new physics searches. Additionally, understanding inelastic neutrino-nucleus scattering is also vital for the detection of core-collapse supernova signals by the next generation neutrino experiments, such as DUNE~\cite{DUNE:2020zfm} and Hyper-K~\cite{Hyper-Kamiokande:2016srs}. At present there are very few existing measurements of inelastic neutrino-nucleus cross sections in this energy regime, leaving them poorly understood.  The measurements that exist are primarily for carbon and iron targets performed by the KARMEN~\cite{Maschuw:1998qh} and LSND~\cite{LSND:2001fbw} collaborations, none at better than the 10\% uncertainty level. Presently no measurement has been made for scattering on the argon nucleus. Theoretical understanding of these processes is also relatively poor, due to strong dependence of the interaction rates on the specific initial- and final-state nuclear wavefunctions and require cumbersome computation to account for underlying complex nuclear structure. It is therefore important to have good theoretical estimates of their rates employing fast and efficient computational methods.\\

A number of theoretical approaches have been used in the past to calculate inelastic neutrino-nucleus scattering~\cite{Haxton:1987kc,Kolbe:1992xu,Ormand:1994js, Kolbe:1999vc, Hayes:1999ew,Volpe:2000zn, Engel:2002hg, Suzuki:2006qd,Suzuki:2009zzc, Cheoun:2011hj,Kostensalo:2018kgh} although with more attention paid to the charged current processes. Additionally, not much attention has been paid to nuclei that are relevant to the current low-energy neutrino experiments, e.g., argon and cesium. More recent studies of neutral current inelastic processes have been presented, e.g., within the continuum Random Phase Approximation (CRPA) method~\cite{VanDessel:2019atx, VanDessel:2020epd}, the deformed shell model mixed with the free nucleon approximation \cite{Sahu:2020kwh}, and merely free nucleon approximation \cite{Bednyakov:2018PRD}. 
The CRPA model employs long-range correlations among nucleons on top of a Hartree-Fock picture of the nucleus and predicts cross sections above the nucleon emission threshold utilizing a multipole expansion~\cite{Nikolakopoulos:2020alk, Pandey:2014tza}. This method to date provides predictions of inclusive cross sections, not cross sections for specific final states which are required for predicting detailed experimental signatures.
The free nucleon approximation is particularly inaccurate in the calculation of inelastic scattering in this energy range since it ignores nuclear structure. While Ref.~\cite{Sahu:2020kwh} does employ the deformed shell model to include nuclear structure effects, their results only consider the first excited state. In this work we provide the first comprehensive calculation of neutral current (NC) inelastic neutrino-nucleus scattering using electroweak theory and the nuclear shell model, and extend it to describe inelastic DM-nucleus scattering in a consistent manner.

Low-energy beam based neutrino experiments (CCM, COHERENT, etc.) provide a great opportunity to probe well-motivated light dark matter via light mediators, where the timing information is utilized to control the SM neutrino background.  The high-intensity proton beam impacts a target producing a high-intensity flux of photons from cascades, meson decays and pion absorption~\cite{Dutta:2020vop}. These photons can then produce light vector mediators via kinetic mixing, while scalar mediators can be produced from three-body decays of charged pions~\cite{Dutta:2021cip}. The light mediators decay promptly into a pair of DM particles which can then produce signals via nuclear recoils in the detector. For proton beams with $E_p\sim1$~GeV the resulting energy of light DM particles is $\sim$ $\mathrm{O}$(50-100)~MeV. Like the neutrinos, the DM particles will also initiate inelastic DM-nucleus scattering. Here, the inelastic scatters involve relativistic DM unlike the non-relativistic DM-nucleus scattering in direct detection experiments, previously explored in~\cite{Baudis:2013bba,
Vietze:2014vsa}. 
In this paper we evaluate these cross sections fully relativistically using electroweak multipole operators, with special attention to the long-wavelength limit Gamow Teller operator.

The DM-nucleus inelastic scattering process is mediated by a neutral dark photon only and thus the NC neutrino-nucleus scattering process will be an important and irreducible background for any accelerator based DM searches using this channel.

The paper is organized as follows, in section II we describe neutrino-nucleus scattering within the electroweak multipole framework, in section III we extend this to the case of DM-nucleus scattering, in section IV we present the shell-model calculation results for the cross sections, in section V, we discuss scattering rates and experimental signatures and lastly, in section VI we conclude.

\section{Neutrino-nucleus scattering}
\subsection{Electroweak multipole operators}
\begin{figure}[ht]
    \includegraphics[width=0.8\columnwidth]{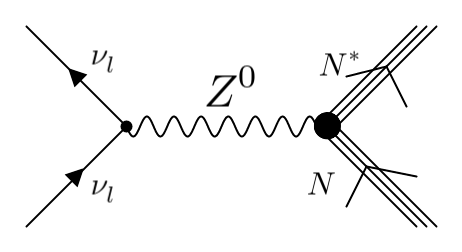}\\
    \caption{The inelastic neutral current neutrino-nucleus scattering process where a $Z^{0}$ boson is exchanged between neutrino ($\nu_{l}$) and the target nucleus ($N$).}
\label{fig:nudiag}
\end{figure}

At sufficiently small momentum transfer (where $q<100$~MeV), neutrinos coherently scatter from  nuclei which, in the case of elastic scattering, greatly enhances the cross section. As shown in Fig.~\ref{fig:nudiag}, the scattering process involves the exchange of a $Z$ boson, where the outgoing nucleus $N^*$ could be in the ground state (for elastic scattering) or an excited state (for inelastic scattering). Inelastic scattering can also proceed via a charged current interaction, in which case the outgoing nucleus will have a different atomic number. In this work we will only consider neutral current scattering.

The differential cross section for the CE$\nu$NS process is conventionally calculated assuming that the protons and neutrons are distributed equally within the nucleus, it then takes the form:
\begin{eqnarray}
    \frac{d\sigma^\nu_{\rm el}}{dE_r} & = &\frac{G_F^2}{4\pi}m_N \left(1-\frac{E_r}{E_\nu}-\frac{m_N E_r}{2E_\nu^2}\right)\nonumber\\
    & & \times\left[(1-4 \sin^2\theta_W)Z-N\right]^2 F_W^2(q^2)
\end{eqnarray}
where $m_N$ is the target nucleus mass, $E_\nu$ is the incoming neutrino energy, $E_r$ is the nuclear recoil energy, $Z$ ($N$) is the target's atomic (neutron) number, $F_W(q^2)$ is the weak form factor of the nucleus, and $\theta_W$ is the Weinberg angle.  A common parameterization for the form factor is the Helm form~\cite{Helm:1956zz}. Given the value of $\sin^2\theta_W \approx 0.23$, the CE$\nu$NS cross section scales approximately as the number of neutrons squared in the target nucleus, and (not including small radiative corrections) is independent of the neutrino flavor. Efforts to improve the CE$\nu$NS cross section calculation through a more detailed treatment of the hadronic current which allows for sub-leading nuclear structure effects was included in Ref.~\cite{Hoferichter:2020osn} while radiative corrections were added in Ref.~\cite{Tomalak:2020zfh}.

In the context of low-energy CE$\nu$NS experiments, the inelastic scattering of neutrinos has not received much attention - potentially owing to its smaller cross section which is not coherently enhanced. Previous attempts to estimate the inelastic neutrino-nucleus scattering rates have not used detailed nuclear calculations~\cite{Bednyakov:2018PRD} or only explored the lowest lying states~\cite{Sahu:2020kwh}. In this work we apply the formalism of semi-leptonic electroweak theory as developed in \cite{DeForest:1966ycn,Serot:1978vj,Donnelly:1979ezn} to the calculation of inelastic neutrino-nucleus scattering. In this formalism the relevant hadronic current is spherically decomposed and expanded in multipoles to obtain irreducible tensor operators which act on single particle states, which can generically be expressed as an expansion of harmonic oscillator states~\cite{Donnelly:1979ezn}. This is a convenient basis to work in, since it results in single-particle matrix elements that are polynomials of the momentum transfer. Following \cite{walecka2004theoretical}, in the extreme relativistic limit the total cross section can be written as:
\begin{widetext}
\begin{align}
    \left(d\sigma \over d\Omega\right)_{\nu,\bar{\nu}} &= \frac{2 G_F^2}{\pi(2J_i+1)} E_f^2 \,\cos^2{{\theta \over 2}} \Bigg\{ \sum_{J=0}^\infty |\langle J_f || \hat{\mathcal{M}}_J + \frac{q_0}{q} \hat{\mathcal{L}}_J || J_i \rangle|^2  \nonumber \\
    &+ \left[-{q_\mu^2 \over 2 q^2} + \tan^2{{\theta \over 2}} \right] \sum_{J=1}^\infty \left[ | \langle J_f || \hat{\mathcal{T}}_J^{el} || J_i \rangle|^2 + |\langle J_f || \hat{\mathcal{T}}_J^{mag} || J_i \rangle |^2 \right]  \nonumber \\
    &\mp 2 \tan{{\theta \over 2}} \left[- {q_\mu^2 \over q^2} + \tan^2{{\theta \over 2}} \right]^{1/2} \sum_{J=1}^\infty {\rm Re} \left(\langle J_f ||\hat{\mathcal{T}}_J^{mag} || J_i \rangle \langle J_f || \hat{\mathcal{T}}_J^{el} || J_i\rangle^* \right) \Biggr\}
    \label{xsec}
\end{align}
\end{widetext}
where $J_{i/f}$ is the initial/final nuclear spin, $E_f = E_\nu - \omega$ is the outgoing neutrino energy, $\theta$ is the scattering angle, $\omega = \Delta E + E_r \approx \Delta E$(excitation energy), $q^\mu=(q_0, \textbf{q})$ is the 4-momentum transfer with $q_0=E_\nu - E_f$ and $q=|\textbf{q}|=\sqrt{2m_N E_r}$ (for a nuclear recoil energy of $E_r$). The electroweak multipole operators are projections of the weak hadronic current, $\hat{\mathcal{J}}$, such that they can act on nuclear states with good angular momentum and parity, they are defined by
\begin{eqnarray}
    \hat{\mathcal{M}}_{JM} &=& \hat{M}_{JM} + \hat{M}^5_{JM} \nonumber\\
        &=& \int d^3x [j_J(qx)Y_{JM}(\Omega_x)] \hat{\mathcal{J}}_0(x) \nonumber\\
    \hat{\mathcal{L}}_{JM} &=& \hat{L}_{JM} + \hat{L}_{JM}^5 \nonumber\\
        &=& \frac{i}{q} \int d^3x \left[ \nabla[j_J(qx)Y_{JM}(\Omega_x)] \right]\cdot \hat{\mathcal{J}}(x) \nonumber\\
    \hat{\mathcal{T}}_{JM}^{\rm el} &=& \hat{T}^{\rm el}_{JM}+\hat{T}_{JM}^{\rm el5}\nonumber\\
        &=& \frac{1}{q} \int d^3x [\nabla \times j_J(qx) \textbf{Y}^M_{JJ1}(\Omega_x)]\cdot \hat{\mathcal{J}}(x)\nonumber\\
    \hat{\mathcal{T}}_{JM}^{\rm mag} &=& \hat{T}_{JM}^{\rm mag} + \hat{T}_{JM}^{\rm mag5} \nonumber\\
        &=& \int d^3x [j_J(qx) \textbf{Y}^M_{JJ1}(\Omega_x)]\cdot \hat{\mathcal{J}}(x)
    \label{eq:mops}
\end{eqnarray}
where $j_J(qx)$ are Bessel functions, $Y_{JM}(\Omega_x)$ are spherical harmonics and $\textbf{Y}^M_{JJ1}(\Omega_x)$ are vector spherical harmonics. The weak hadronic current has V-A structure $\hat{\mathcal{J}}_\mu=\hat{J}_\mu + \hat{J}_\mu^5$, allowing the operators to be split into components of normal and abnormal parity. The normal parity operators: $M_{JM}, L_{JM}, T^{el}_{JM},$ and $T_{JM}^{mag5}$ can only contribute to transitions with $\Delta \pi = (-1)^J$. Similarly only $M^5_{JM}, L^5_{JM}, T^{el5}_{JM},$ and $T_{JM}^{mag}$ can contribute to abnormal parity transitions $\Delta \pi = (-1)^{J+1}$. 
At the one-body level, the vector and axial multipole operators can be expressed in terms of the 7 single-particle operators and nucleon form factors (see \cite{Haxton:2008zza} for further details):
\begin{eqnarray}
    M_{JM} &=& F_1^{N} (q_\mu^2) M_J^{M}\nonumber\\
    L_{JM} &=& \frac{q_0}{q} M_{JM}\nonumber\\
    T^{\rm el}_{JM}&=& 
    {q \over m_n} (F_1^{N} (q_\mu^2) \Delta^{'M}_J + {\textstyle {1 \over 2}} \mu^{N}(q_\mu^2) \Sigma^{M}_J)\nonumber\\
    T^{\rm mag}_{JM}&=&
    -{iq \over m_n} (F_1^{N}(q_\mu^2) \Delta_J^{M} - {\textstyle {1 \over 2}} \mu^{N}(q_\mu^2) \Sigma_J^{'M})\nonumber\\
    M^{5}_{JM} &=&
    {iq \over m_n} (G_A^{N}(q_\mu^2) \Omega_J^{'M} + {\textstyle {1 \over 2}} q_0 G_P^{N}(q_\mu^2) \Sigma^{''M}_J)\nonumber\\
    L_{JM}^5 &=& 
    i\left(G_A^{N}(q_\mu^2)-{q^2 \over 2m_n} G_P^{N}(q_\mu^2) \right) \Sigma^{''M}_J \nonumber\\
    T_{JM}^{\rm el5} &=& \label{eq:Tel5} iG_A^{N}(q_\mu^2) \Sigma^{'M}_J\nonumber\\
    T_{JM}^{\rm mag5} &=& G_A^{N}(q_\mu^2) \Sigma^{M}_J
\end{eqnarray}
where $m_n$ is the nucleon mass, $\mu^{N}(q_\mu^2) = F_1^{N}(q_\mu^2) + 2m_n F_2^{N}(q_\mu^2)$ is the magnetic moment, and $F_1(q_\mu^2)$, $F_2(q_\mu^2)$, $G_A(q_\mu^2)$ and $G_P(q_\mu^2)$ are the Dirac, Pauli, axial and pseudoscalar neutral current nucleon form factors. In the low recoil-energy limit, these become~\cite{Hoferichter:2020osn}: 
\begin{eqnarray}
    F_1^{n}(0)=&0      \;\;\;\;  &F_1^{p}(0)=1 \nonumber\\
    F_2^{n}(0)=&-1.91  \;\;\;\;  &F_2^{p}(0)=1.79 \nonumber\\
    G_A^{n}(0)=&-g_A/2 \;\;\;\;  &G_A^{p}(0)=g_A/2 \nonumber\\
    G_P^{n}(0)=&{2m_n G_A^{n}(0) \over m_\pi^2} \;\;\;\;  &G_P^{p}(0)= {2 m_n G_A^{p}(0) \over m_\pi^2} \nonumber
\end{eqnarray}

The energy dependence of the axial form factor $G_A$ is conventionally taken to have a dipole form \cite{Bernard_2001},
\begin{equation}
    G_A(q^2)=\frac{g_A}{(1+\,q^2/\Lambda_A^2)^2}
\end{equation}
where $g_A=1.27$ and the axial mass is $\Lambda_A \approx 1040$~MeV. In this work the scattering processes we consider have low momentum transfer, $q^2 \ll \Lambda_A^2$, and thus the $q^2$ dependence of $F_A$ can be neglected.

The results of the the single particle operators acting on harmonic oscillator basis states have been tabulated in~\cite{Donnelly:1980tsp}. In this work we use the SevenOperators code to evaluate and simplify the relevant matrix elements~\cite{Haxton:2008zza}. Sometimes it is desirable to have the differential cross section in Eq.~(\ref{xsec}) written in terms of recoil energy, i.e. $\frac{d\sigma}{dE_r}=\frac{d\sigma}{d\Omega} \frac{d\Omega}{dE_r}=2\pi\frac{d\sigma}{d\Omega}\frac{m_N}{E_i E_f}$. 

In this analysis we consider only one-body currents, however, there are likely significant contributions from two-body currents~\cite{Hoferichter:2020osn}.

\subsection{Gamow–Teller transitions}
\label{sec:GT}
For small momentum transfers, when a Gamow-Teller (GT) transition is kinematically accessible, the inelastic cross section will be dominated by such allowed GT transitions. We can see this by looking at Eq.~(\ref{eq:mops}) in the long-wavelength limit. The only surviving multipoles are $\hat{\mathcal{M}}_{00}$, $\hat{\mathcal{T}}^{el}_{1M}$ and $\hat{\mathcal{L}}_{1M}$. With $\hat{\mathcal{L}}$ further suppressed by $q$ in the cross section (see Eq.~(\ref{xsec})), we need only consider the first two operators. The $\hat{\mathcal{M}}_{00}$ and $\hat{\mathcal{T}}^{el}_{1M}$ operators are associated with Fermi operator, $\hat{\tau_0}$, and GT operator, $\frac{1}{2}\hat{\sigma_i} \hat{\tau_0}$, respectively: 
\begin{eqnarray}
    \hat{\mathcal{M}}_{00} &=& \frac{1}{\sqrt{4\pi}} F_1\,\, \sum_{i=1}^A \hat{FT}\\
    \hat{\mathcal{T}}^{el}_{1M} &=& \sqrt{2} \hat{\mathcal{L}}_{1M} = \frac{i}{\sqrt{6\pi}}G_A\,\, \sum_{i=1}^A \hat{GT}.
    \label{eq:GTFT}
\end{eqnarray}
In the long-wavelength limit the GT operator allows transitions with $|\Delta J|=1$, while the Fermi operator allows $\Delta J=0$ transitions. For inelastic scattering, $\Delta J=0$ transitions are possible only if a nucleon can be excited to a state of the same $l$ but higher $n$ (since conserving $J$ but changing $l$ by 1 is disallowed by parity conservation). For the model spaces we consider such a transition does not exist. So while such a possibility is allowed by our multipole calculation, it does not contribute here.

Therefore, considering the GT operator alone in the long wavelength limit can provide an efficient way to approximate the scattering cross section. In the multipole operator analysis the GT operator is contained within $\hat{T}_J^{el5}$. The relevant part of the cross section in Eq.~(\ref{xsec}) is:
\begin{align}
    \label{eq:xsecTel5}
    {{d\sigma^{\rm GT} \over d\Omega}} & =  \frac{2 G_F^2}{\pi(2J+1)} E_f^2 \,\cos^2{{\theta \over 2}} \nonumber \\
    & \times \left(-{q_\mu^2 \over 2 q^2} + \tan^2{{\theta \over 2}} \right) \sum_{J=1}^\infty | \langle J_f || \hat{T}_J^{el5} || J_i \rangle|^2. 
\end{align}
Taking the long–wavelength limit ($q\rightarrow 0$, or equivalently $E_r\rightarrow 0$) gives $G_A \approx g_A$ and ignores the momentum dependence of the nuclear form factor, the amplitude can then be simplified to:
 \begin{equation}
    \bigl| \langle J_f ||\sum_{J=1}^\infty \hat{T}_J^{el5} || J_i \rangle \bigr|^2 \approx \frac{g_A^2}{6\pi} \bigl| \langle J_f || \sum_{i=1}^A \frac{1}{2}\hat{\sigma_i} \hat{\tau_0} || J_i \rangle \bigr|^2.
\end{equation}
Substituting this into Eq.~(\ref{eq:xsecTel5}) allows us to write the cross section as
\begin{align}
\label{eq:GTnuAngle}
    {d\sigma^{\rm GT} \over \ d \cos\theta} &\approx \frac{G_F^2 g_A^2}{2\pi (2J+1)} (E_\nu - \Delta E)^2 \left( 1-\frac{1}{3}\, \cos\theta \right)\\
    & \times |\langle J_f|| \sum_{i=1}^A \frac{1}{2}\hat{\sigma_i} \hat{\tau_0}|| J_i\rangle|^2 \nonumber
\end{align}
where we have written the final neutrino energy in terms of the incoming neutrino energy, $E_\nu$, and the final state excitation energy, $\Delta E$, which is assumed to be small in this approximation. We can then make connection with the common form of the cross section for neutral current GT transitions by integrating Eq.~(\ref{eq:GTnuAngle}) to find the total cross section: 
\begin{equation}
    \sigma^{GT}_\nu \approx \frac{G_f^2 g_A^2}{\pi (2J+1)} (E_\nu- \Delta E)^2 |\langle J_f|| \sum_{i=1}^A \frac{1}{2}\hat{\sigma_i} \hat{\tau_0}|| J_i\rangle|^2.
    \label{eq:GTnu}
\end{equation}
This form of the cross section is in agreement with others found in the literature~\cite{hnps2493,Gardiner:PRC2021}. We will see in a later section that using Eq.~(\ref{eq:GTnu}) to calculate the cross section will allow for an important simplification of the numerical many-body problem while providing an adequate estimation of the transition matrix elements in the long wavelength limit.

\section{Dark matter-nucleus scattering}
In this section we apply the formalism discussed in the preceding section to the inelastic scattering of dark matter (DM) on nuclei. Previous analyses considered the scattering of nonrelativistic dark matter \cite{Sahu:2020kwh,Klos:PRD2013,Arcadi:2019hrw}, which greatly restricts the excited states that are accessible.  Astrophysically, dark matter can be boosted via several mechanisms, however the flux will be subdominant compared to the non-relativistic component. Larger fluxes of boosted dark matter could be produced in the decay of particles produced in collider experiments such as \cite{Akimov:2017ade,CCM}. 

\begin{figure}[t]
    \centering
    \includegraphics[width=0.8\columnwidth]{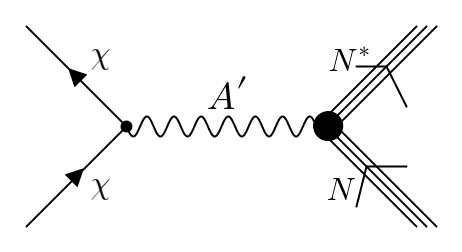}
    \caption{Scattering of a DM particle, $\chi$, from a nucleus, $N$, mediated via a dark photon, $A^\prime$.}
    \label{fig:dmdiag}
\end{figure}

Low-energy beam dump experiments can investigate light dark matter where the light dark matter interacts with the SM particle via light mediators. In these experiments neutrinos are produced as a product of stopped-pion decay, but they are also high-intensity sources of photons emerging from meson decays and cascades which could produce exotic light vector mediators via kinetic mixing. The light vector mediators can then promptly decay into a pair of dark matter particles ($\chi,\bar{\chi}$) which are semirelativistic. There exist many well motivated models where this scenario is possible~\cite{Huh:2007zw,Pospelov:2007mp,Chun:2010ve,Batell:2014yra,
deNiverville:2015mwa,Kaplan:2009ag,Bi:2009uj,Kim:2015fpa, Foldenauer:2018zrz,Dutta:2019fxn}. As a benchmark example we will consider a dark photon A-prime ($A'$) model where the $A'$ undergoes kinetic mixing with the SM photon. The model is described by the Lagrangian:
\begin{equation}
    \mathcal{L} \supset g_D A'_\mu \bar \chi \gamma^\mu \chi + e \epsilon Q_q A'_\mu \bar q \gamma^\mu q
\end{equation}
where $g_D$ is the dark coupling constant, $\epsilon$ is the mixing parameter, $Q_q$ is quark's electric charge. The dark photon will be produced in the processes of pion capture, pion decay and the photons emerging from the cascades:
\begin{eqnarray}
    \pi^- + p &\rightarrow& n + A' \\
    \pi^+ + n &\rightarrow& p + A' \\
    \pi^0 &\rightarrow& \gamma + A'
\end{eqnarray}
Via these processes the SNS, for example, produces dark photons at a rate of $\epsilon^2 \times 0.23 \times 10^{20}$ per day, mostly from $\pi^0$ decay. The dark photons then decay to DM: $A' \rightarrow \chi\bar{\chi}$. Previous analyses looked at the elastic scattering of such DM from nuclei in the COHERENT detectors \cite{Dutta:2020vop}, here we extend upon this to include inelastic scattering. 

The scattering of DM from nuclei proceeds via the exchange of a dark photon, as depicted in Fig.~\ref{fig:dmdiag}. The cross section of elastic DM-nucleus scattering is given by \cite{Dutta:2020prl}:
\begin{eqnarray}
    \frac{d\sigma^{DM}_{\rm el}}{dE_r} &= \frac{e^2\epsilon^2 g_D^2 Z^2}{4\pi(E_\chi^2 - m_\chi^2)(2m_N E_r + m_{A'}^2)^2} F^2(q^2) \nonumber\\
    \times&\left[
    2E_\chi^2 m_N \left(1-\frac{E_r}{E_\chi}-\frac{m_N E_r}{2 E_\chi^2} \right) + E_r^2 m_N
    \right]
    \label{eq:DMel}
\end{eqnarray}
where $E_\chi/p_\chi$ is the energy/momentum of the incoming DM, and $F(q^2)$ denotes the nuclear form factor. \footnote{The form factor $F$ refers to the elastic charge form factor of the nucleus, for simplicity we take it to be of the Helm form.} We assume the $A'$ decays promptly into a pair of DM particles in this work.

Following the multipole formalism discussed in the previous section, the inelastic cross section is found to be
\begin{widetext}
\begin{equation}
\label{eq:dmInelCS}
\begin{split}
    \frac{d\sigma^{DM}_{\rm inel}}{dE_r}& = \frac{2e^2\epsilon^2 g_D^2 {E'}_\chi^2}{p_\chi {p'}_\chi(2m_N E_r + m_{A'}^2 )^2} \frac{m_N}{2\pi} \frac{4\pi}{2J+1} \Bigg\{
    \sum_{J\geqslant 1, spin} \left[ \frac{1}{2}(\vec{l} \cdot \vec{l}^* - l_3 l_3^*) \left(
        \bigl| \langle J_f ||\hat{\mathcal T}_J^{mag}||J_i \rangle \bigr|^2 + \bigl| \langle J_f ||\hat{\mathcal T}_J^{el}|| J_i \rangle \bigr|^2
    \right) \right] \\
    & +\sum_{J\geqslant 0, spin} \left[ l_0 l_0^* \, \bigl| \langle J_f ||\hat{\mathcal M}_J|| J_i\rangle \bigr|^2 + l_3 l_3^* \, \bigl| \langle J_f||\hat{\mathcal L}_J|| J_i\rangle \bigr|^2 - 2\, l_3 l_0^* Re\left( \, \langle J_f||\hat{\mathcal L}_J|| J_i\rangle \langle J_f ||\hat{\mathcal M}_J|| J_i \rangle ^*
    \right) \right]\Bigg \}
\end{split}
\end{equation}
\end{widetext}

where $E_\chi'$ and $p_\chi'$ are the outgoing DM energy and momentum. While the neutrino current contained an axial component, the DM current is purely vector: $l_\mu=\bar \chi \gamma^\mu \chi$. The dark matter current terms $\sum\limits_{s_i,s_f} l_\mu l_\nu^*$ of Eq.~(\ref{eq:dmInelCS}) evaluate to:
\begin{align}
    &\sum\limits_{s_i,s_f} l_0 l_0^* = 1 + \frac{1}{4E_\chi {E'}_\chi} \left(2 p_\chi^2 +2{p'}_\chi^2-4 m_N E_r+m_\chi^2 \right) \nonumber\\
    &\sum\limits_{s_i,s_f} l_3 l_3^* = 1+ \frac{1}{E_\chi {E'}_\chi} \left[ -{3\over2} \left(p_\chi^2 +{p'}_\chi^2 \right) +3m_NE_r - {m_\chi^2\over4} \right] \nonumber\\
    &\sum\limits_{s_i,s_f} l_3 l_0^* = -\left({ p_\chi^2 + {p'}_\chi^2-2m_NE_r \over 2p_\chi E_\chi}+{p_\chi \over E_\chi}\right) \nonumber\\
    &\sum\limits_{s_i,s_f} \vec{l} \cdot \vec{l}^* = 3- {1\over E_\chi {E'}_\chi} \left[ {1\over2} \left(p_\chi^2 + {p'}_\chi^2-2m_NE_r \right) +{3m_\chi^2\over4} \right] \nonumber\\
    &\sum\limits_{s_i,s_f} (\vec{l} \times \vec{l}^*)_3 = 0
\end{align}
Further details of the current calculation is given in appendix~\ref{app:dmcurrent}.

As in the case of neutrino-nucleus scattering, the elastic cross section receives a coherent enhancement, in this case by a factor of $Z^2$ because the dark photon couples to the electromagnetic current. Therefore, the ratio of elastic to inelastic events will be larger in light nuclei (e.g. Ar) and smaller in heavy nuclei (e.g. Cs and I). This is the reverse of the neutrino case (which dominantly couple to neutrons); a feature that could aid in differentiating a dark matter signal from the neutrino background.

Similar to neutrino scattering, we can write down the differential cross section for inelastic DM scattering via the GT operator:
\begin{align}
    \label{eq:GTdm}
    \frac{d\sigma^{GT}_{DM}}{dE_r} &= \frac{2e^2\epsilon^2 g_D^2 {E'}_\chi^2}{p_\chi p_\chi'(2m_N E_r + m_{A'}^2 )^2} \frac{m_N}{2\pi} \frac{4\pi}{2J+1} \\
    &\times \biggl[
    \frac{(\vec{l} \cdot \vec{l}^* - l_3 l_3^*)}{2} |\langle J_f|| \hat{\mathcal T}_J^{el5} || J_i\rangle|^2 \nonumber \\
    &+ l_3 l_3^* \, \bigl| \langle J_f||\hat{\mathcal L}_J|| J_i\rangle \bigr|^2 \biggr]. \nonumber
\end{align}
Combining Eq.~(\ref{eq:GTFT}) and  Eq.~(\ref{eq:GTdm}) then gives:
\begin{align}
    \frac{d\sigma^{GT}_{DM}}{dE_r} &= \frac{2e^2\epsilon^2 g_D^2 {E'}_\chi^2}{p_\chi p_\chi'(2m_N E_r + m_{A'}^2 )^2} \frac{m_N}{2\pi} \frac{4\pi}{2J+1} \\
    &\times \frac{\vec{l} \cdot \vec{l}^*}{2} \frac{g_A^2}{6\pi}
    |\langle J_f|| \sum_{i=1}^A \frac{1}{2}\hat{\sigma_i} \hat{\tau_0}|| J_i\rangle|^2. \nonumber
\end{align}

In principle one can integrate the differential cross section over $E_r$ to get the total cross section, however, in this case a closed form expression of Eq.~(\ref{eq:dmInelCS}) could not be obtained.

\section{Shell model calculations}

\subsection{Computational details}
We compute the cross sections of inelastic neutrino/DM-nucleus scattering for $^{40}$Ar, $^{133}$Cs, and $^{127}$I nuclei since these nuclear targets are utilized at the ongoing COHERENT and CCM experiments. Two approaches to the calculation are taken: the full multipole operator analysis and the GT transition in the long wavelength limit. For both approaches we use the nuclear shell model code BIGSTICK \cite{Johnson:2018hrx,Johnson:2013bna} to solve the many-body problem. In general BIGSTICK supports computing not only low-lying states and their density matrices, but also the strength of arbitrary operators with a well-defined basis, e.g. the GT operator. To evaluate the cross sections using the multipole analysis the full nuclear one-body density matrix is required, which BIGSTICK can generate for converged eigenstates. We use one-body density matrices as defined in BIGSTICK:
\begin{equation}
    \rho^{fi}_K(a^\dagger b)= {1\over 2K+1} \langle J_f|| (a^\dagger b)_K || J_i\rangle
\end{equation}
where, following Edmonds and Sakurai, the reduced matrix elements are defined as:
\begin{equation}
    \langle J_f|| O_K || J_i\rangle = {\langle J_f M_f, K M| J_i M_i \rangle^{-1} \over 2J_f+1} \langle J_f M_f| O_K M| J_i M_i\rangle.
\end{equation}
The multipole operator matrix elements can then be computed with the aid of the Mathematica package SevenOperators~\cite{Haxton:2008zza}. Because this approach requires converged eigenstates, it is time and memory consuming if one is interested in many final states. For the restricted case of the GT transition one can use the simpler and more efficient Lanczos strength function~\cite{ca05,bloom1984gamow,BloomFuller1985}, which is directly incorporated into BIGSTICK. After $n$ iterations this method can accurately estimate the $n^{th}$ moment of the distribution. Since we do not require the full convergence of the Lanczos algorithm, only convergence of the integrals over the strengths (i.e. $\int S(E) dE = \sum_{J_f} |\langle J_f|| \hat{O} || J_i\rangle|^2$) this method is highly efficient.

While computing the full density matrix will give more accurate results (including contributions for all operators), we find that the GT strength is sufficiently precise for our purposes. Importantly, when dealing with a large number of final states, it makes the calculation a tractable computational task. While one can perform the strength function calculation for any well-defined operator, we limit ourselves to considering the GT operator which is dominant in the long-wavelength limit. Moreover, it is challenging to explicitly find the basis $\langle a | \hat{O} | b \rangle$ for the other multipole operators.

When computing density matrices we are able to calculate the transition from the ground state to the, for example, $N=2{\,\text-\,}16$ excited states of $^{40}$Ar, which should align well with the lowest lying experimental nuclear energy levels. In comparison, when computing the GT transition with the strength function method, we are able to calculate the strengths of up to $N=500$ states (including states with zero strength). However, since this method doesn't have fully converged states, these states are drawn from a wider range of energies.

For $^{133}$Cs and $^{127}$I, the nucleons are in the orbits $0g_{7/2}$, $1d_{5/2}$, $0h_{11/2}$, $1d_{3/2}$, $2s_{1/2}$ and we use jj55pna interaction \cite{Brown:PRC2001}. $^{40}$Ar has a more challenging nucleon configuration; the protons (10) are in $sd$ orbits ($0d_{5/2}$, $1s_{1/2}$, $0d_{3/2}$), while the neutrons (14) are in $pf$ orbits ($0f_{7/2}$, $1p_{3/2}$, $0f_{5/2}$, $1p_{1/2}$). In that case we consider the nucleons are in $sdpf$ orbits, then perform a truncation to reduce the computational workload. For this model space we apply the SDPF-NR interaction  \cite{Prados:PRC2007,Nowacki:PRC2009,Nummela:PRC2001}. The truncation applied gives higher levels more weight for protons across the $sdpf$ orbits, and limits the maximum number of excited protons which jump to the $pf$ orbits to 4. Neutrons are restricted to the $pf$ orbits.

\subsection{Neutrino-nucleus cross sections}

\begin{figure}[tb]
\centering
\includegraphics[width=0.9\columnwidth]{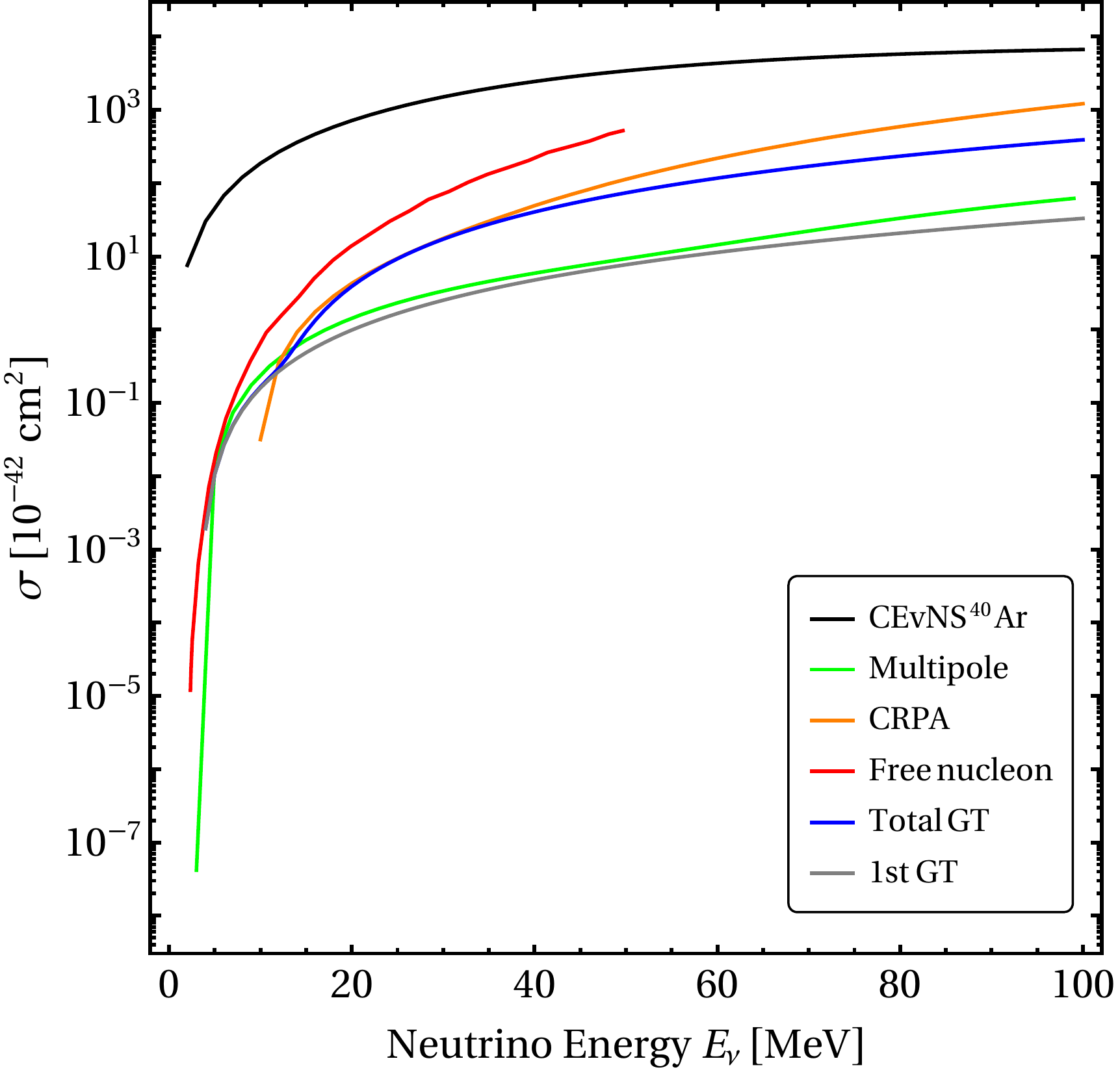} \\
 \includegraphics[width=0.9\columnwidth]{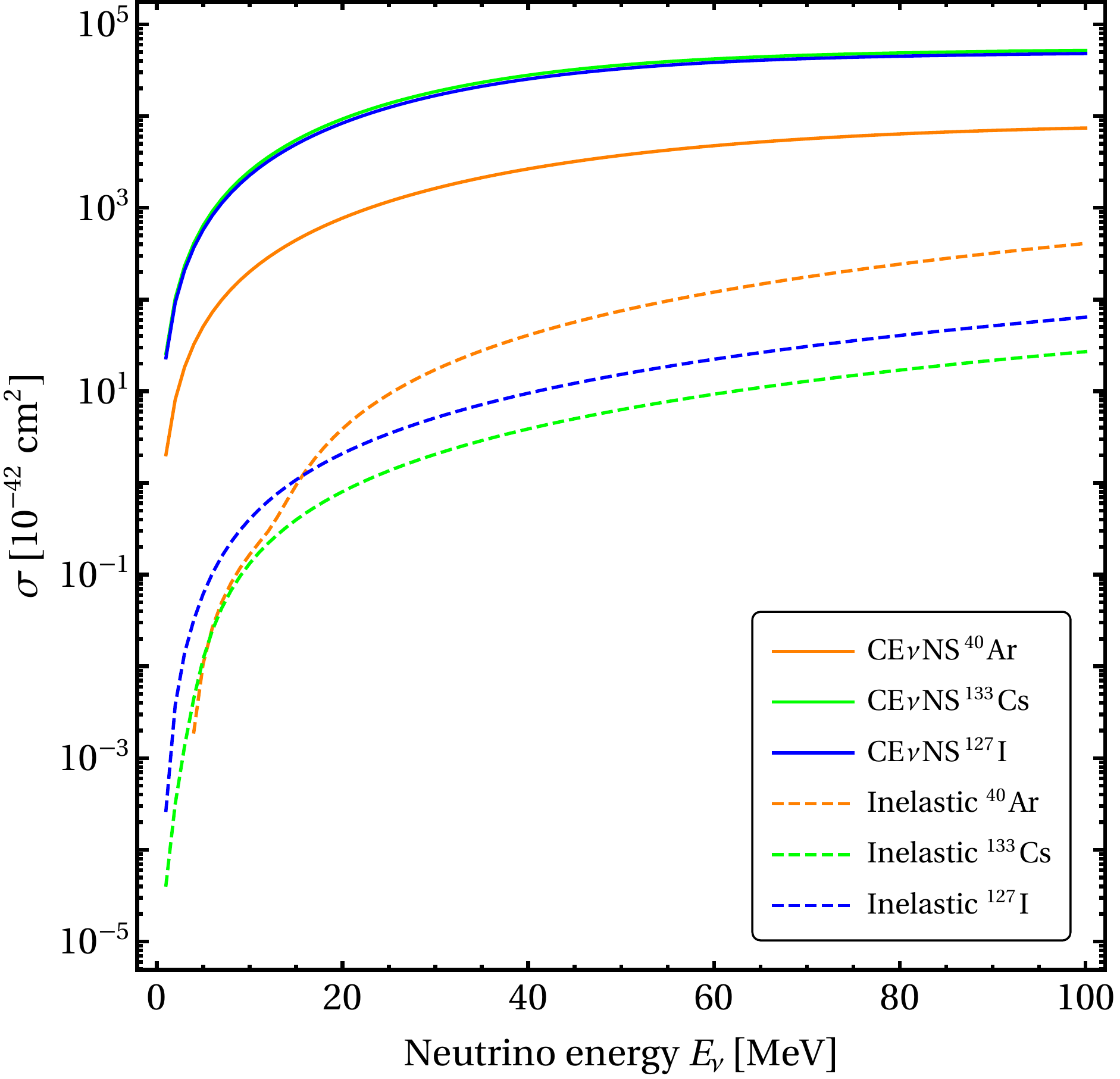}\\
\caption{Top: elastic and NC inelastic $\nu$-$^{40}$Ar cross sections as a function of the incoming neutrino energy. For comparison our calculations (Multipole, Total GT and 1st GT) are compared with the CRPA \cite{VanDessel:2020epd} and free nucleon~\cite{Bednyakov:2018PRD} predictions. Bottom: total elastic and NC inelastic neutrino-nucleus scattering cross sections for $^{40}$Ar, $^{133}$Cs, and $^{127}$I nuclei.}
\label{fig:nuCS}
\end{figure}

In Fig.~\ref{fig:nuCS}, we compare the inelastic cross section result of our two calculations, multipole and GT, with the CE$\nu$NS cross section and with two other inelastic results from the literature. While different calculations show some agreement at lower energies but diverge at higher energy. Our calculation of the total inelastic cross section (given as the sum over all GT transitions) is around an order of magnitude lower than that of Refs.~\cite{Bednyakov:2018PRD} and more closely follows that of~\cite{VanDessel:2020epd} until around $E_\nu=40$~MeV. The former used the free nucleon approximation to calculate just one excited state and so agreement is only expected for very small neutrino energies when few states are kinematically accessible. The latter calculates the inclusive inelastic cross section, above nucleon emission threshold, for low-lying states through the Continuum Random Phase Approximation (CRPA). Working with a limited spectrum of excited states, our multipole results can only include transitions to the first 15 excited states and thus it is unable to capture all accessible transitions and therefore is not an accurate estimate of the total cross section above $E_\nu\gtrsim10$~MeV. However, we can use it to assess the contribution of non-GT transitions since the limited spectrum contains one $J=1$ state ($N=9$) that is accessible via a GT transition. For comparison we have plotted the GT cross section, Eq.~(\ref{eq:GTnu}), for the lowest lying $J=1$ state (gray curve in Fig.~\ref{fig:nuCS} top). This confirms that the GT transitions provide the dominant contribution to the inelastic cross section. Additionally we show the differential cross section as a function of angle and the recoil energy in appendix~\ref{app:shellmodel}.

\begin{figure*}
    \centering
    \includegraphics[width=0.68\columnwidth]{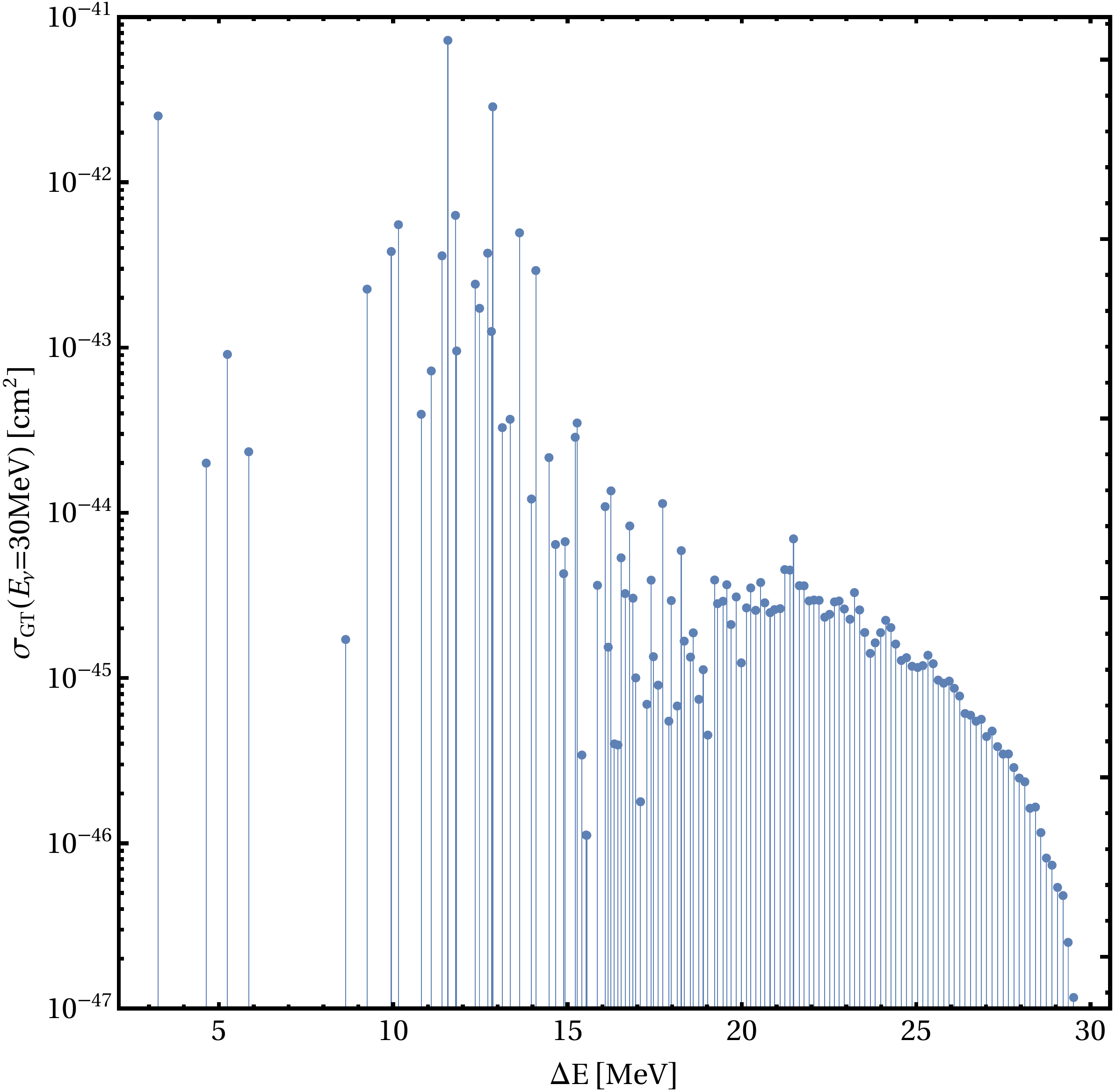}
    \includegraphics[width=0.68\columnwidth]{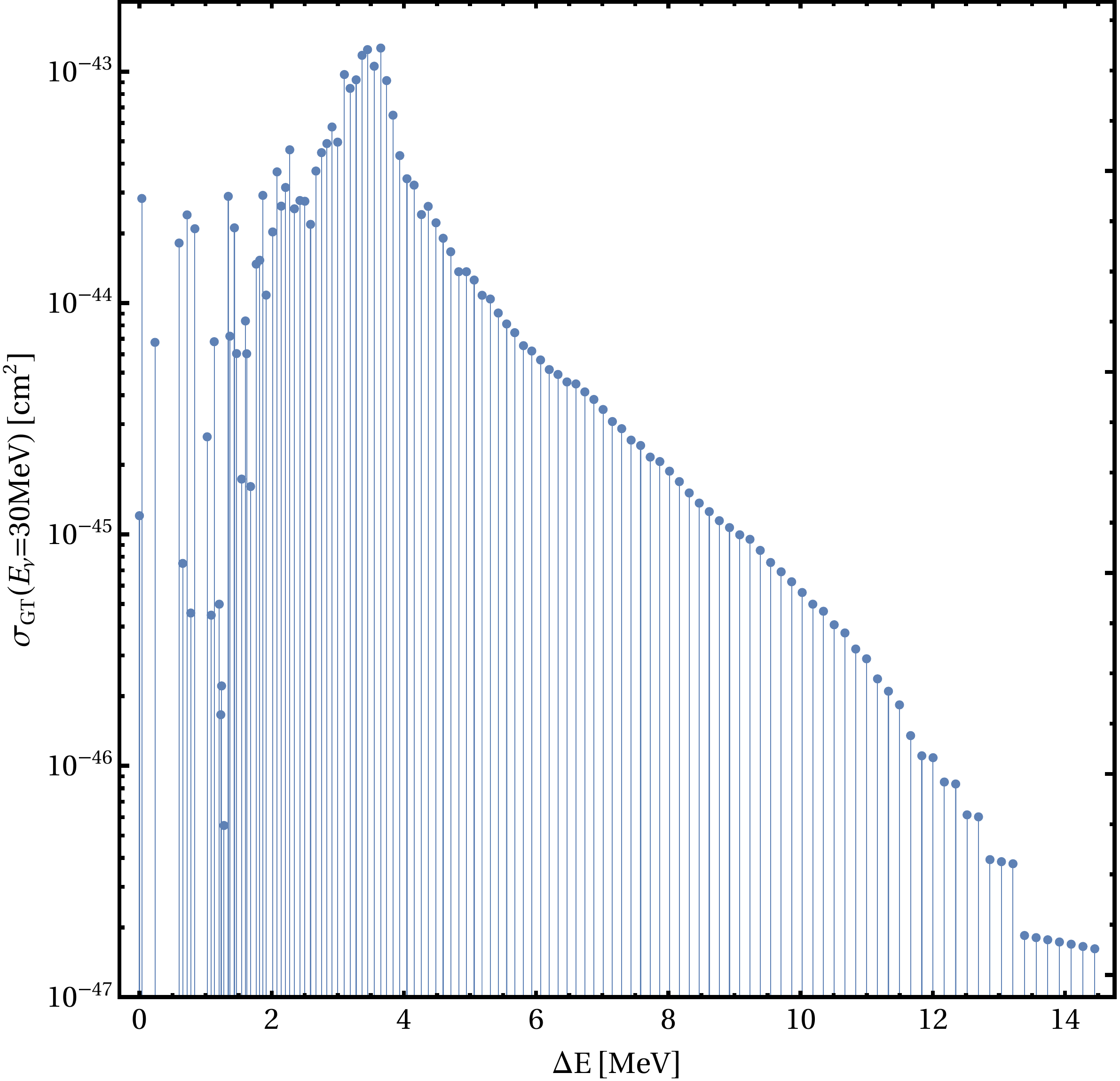}
    \includegraphics[width=0.68\columnwidth]{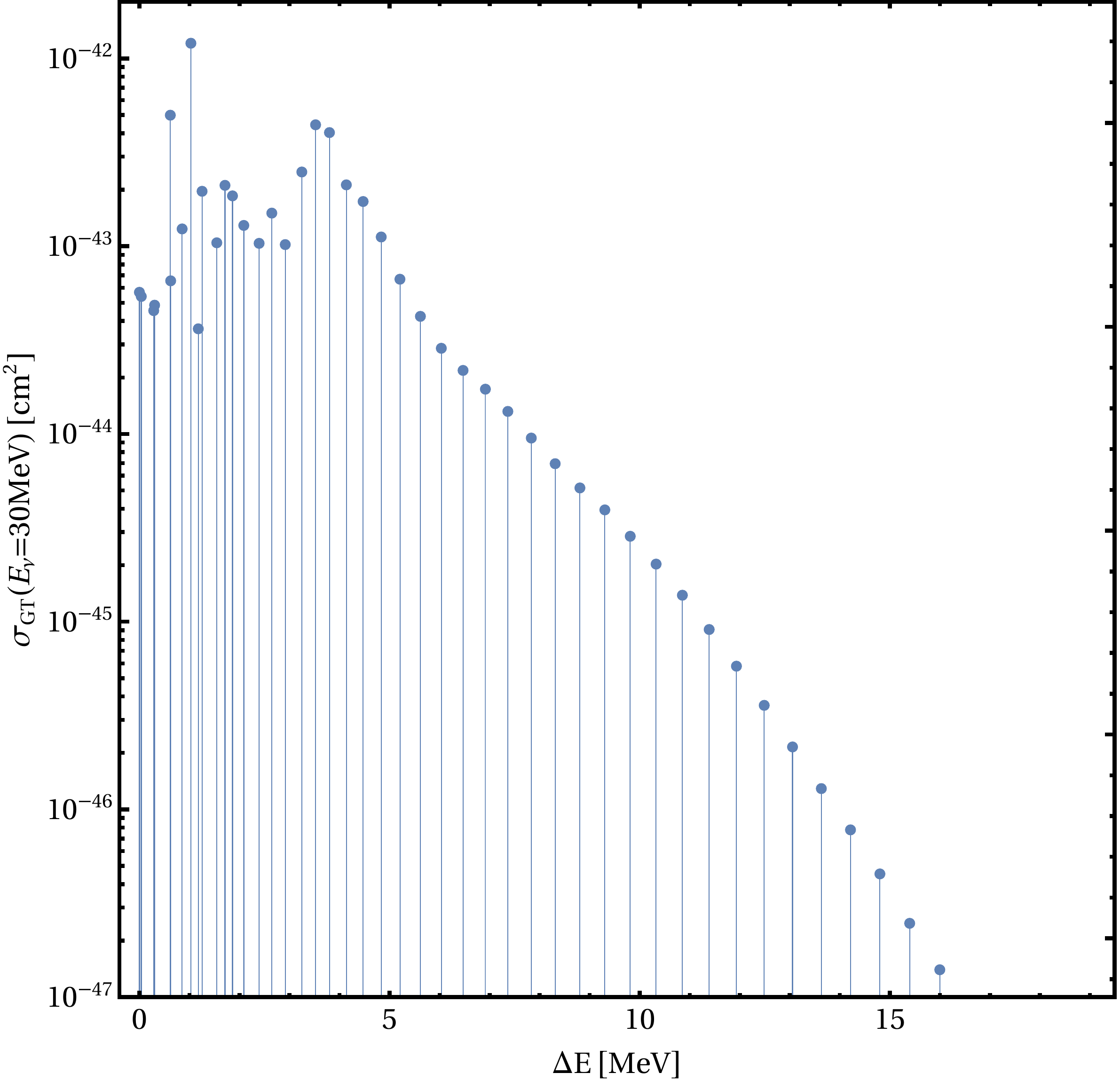}\\
    \includegraphics[width=0.68\columnwidth]{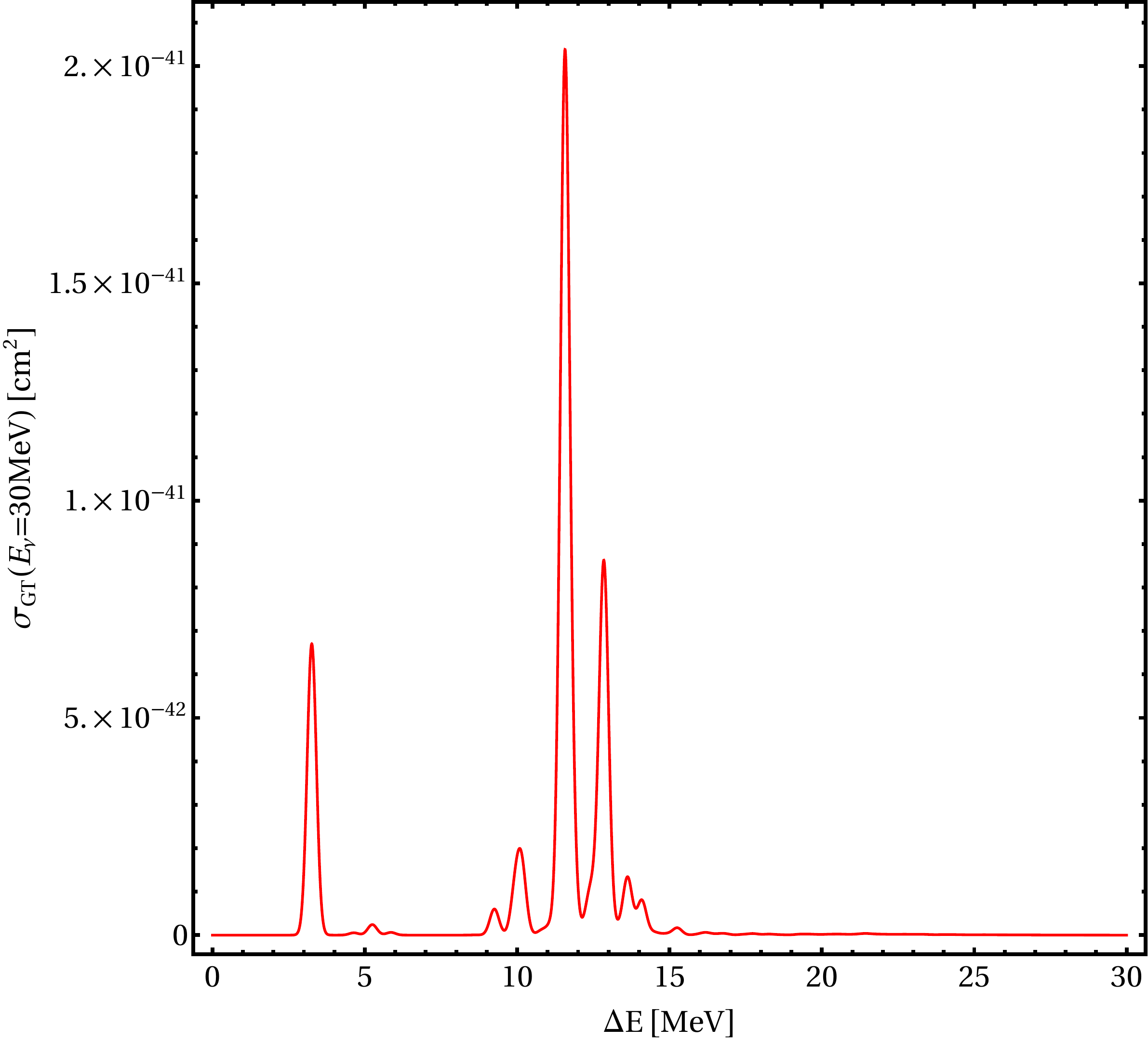}
    \includegraphics[width=0.68\columnwidth]{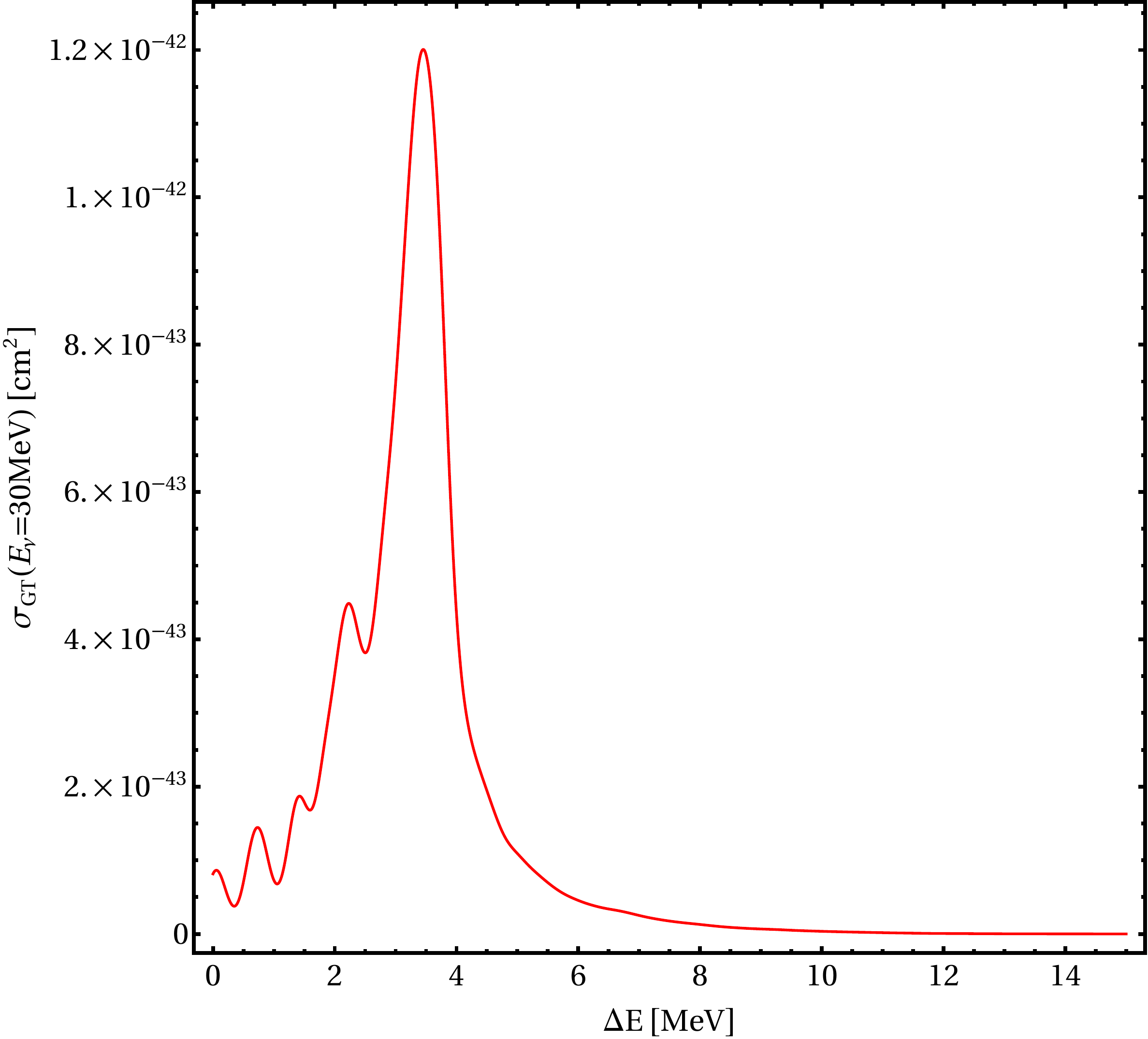}
    \includegraphics[width=0.68\columnwidth]{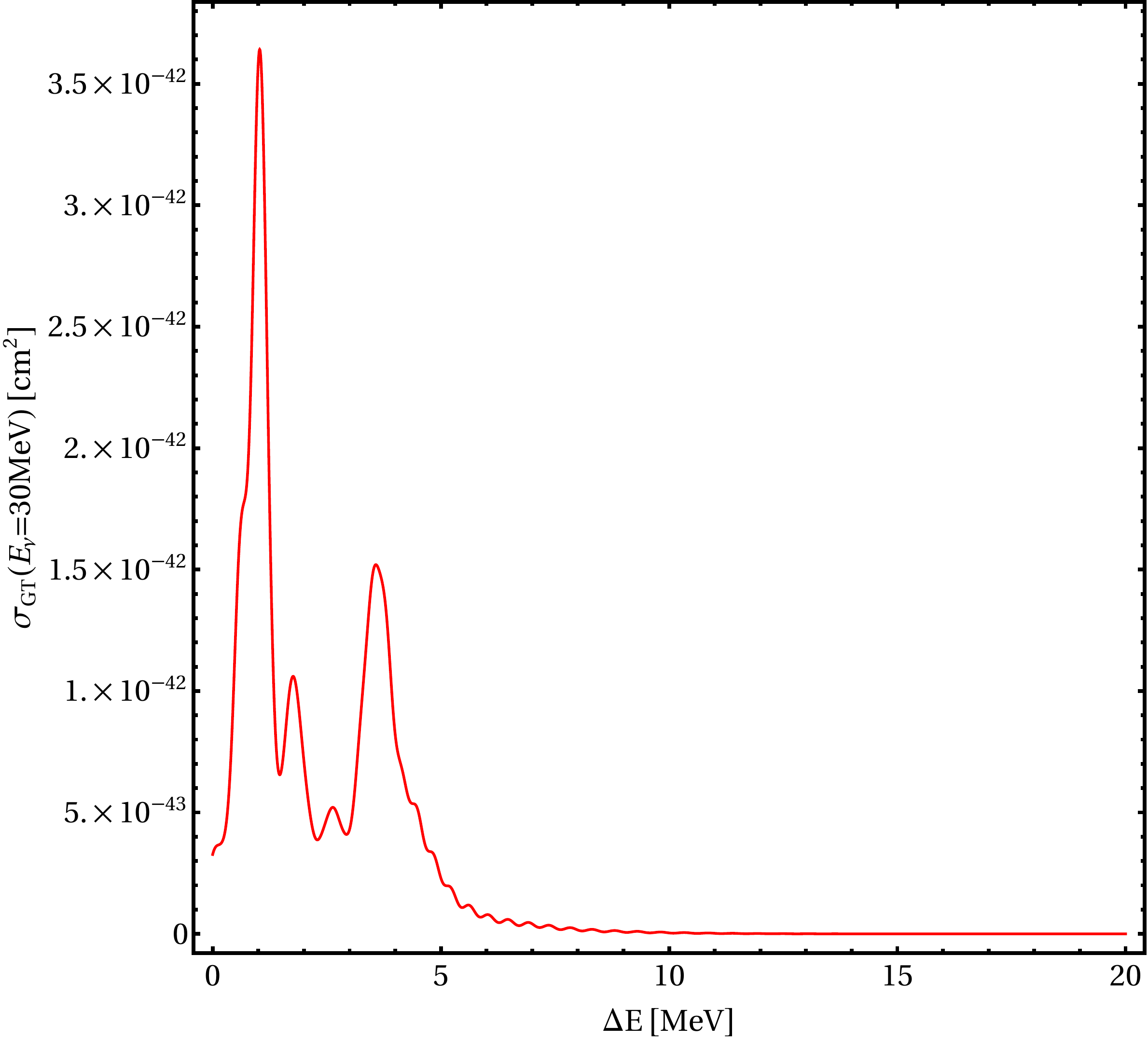}\\
    \caption{The GT transition cross section for each state vs. the excitation energy in $^{40}$Ar (left), $^{133}$Cs (center) and $^{127}$I (right), for incident neutrinos with $E_\nu=30$ MeV. The top row shows the cross sections for each discrete energy level, while the bottom row shows these levels convolved with a Gaussian of width 150 keV.}
    \label{fig:nuGT30}
\end{figure*}

Additional terms in the multipole expansion do contribute and are required for a precise calculation of the cross section for any specific transition. However, when estimating the total cross section, based on the limitations of our calculation, it is much more accurate to ensure that all accessible GT transitions are included. For the same number of states, the GT analysis generally consumes much less computational resources than the multipole analysis. For this reason and the aforementioned improvement in accuracy we use the GT analysis for our results in the following sections.

The total elastic and inelastic cross-sections for the $^{133}$Cs and $^{127}$I targets are shown in Fig.~\ref{fig:nuCS}, where we have included GT transitions to 300 and 100 states (including the states with zero strength) for $^{133}$Cs and $^{127}$I, respectively. The elastic cross sections of both nuclei are virtually the same as they have a similar number of neutrons. In contrast, $^{127}$I has slight higher inelastic cross section since it has a higher total GT strength than $^{133}$Cs, as shown in Fig.~\ref{fig:nuGT30}.

Experimentally the signature of an inelastic collision will be dominated by the deexcitation energy of the nucleus. The excited nucleus falls back to the ground state very quickly through the emission of photons and (if energetically allowed) neutrons. Previous studies focused on the nuclear recoil energy, which is small in comparison, but could in principle be measured (recoil spectra are provided in appendix~\ref{app:shellmodel}). While a full calculation of the observational signatures is beyond the scope of this work, we can use the cross section to each excited state to visualize the relative strength of the energy depositions for each of the allowed transitions. In Fig.~\ref{fig:nuGT30} we plot the cross section for each transition vs. the excitation energy for $^{40}$Ar, $^{133}$Cs and $^{127}$I for a incoming neutrino energy of 30~MeV.

\subsection{DM-nucleus cross sections}
Following the method of the previous section we computed the total cross section for dark matter inelastic scattering using the GT operator in the long wavelength limit. Similar to the neutrino case, we find that GT transitions also dominate the total cross section, as shown in Fig.~\ref{fig:dmGTratio}

\begin{figure}[tbh]
    \centering
    \includegraphics[width=0.9\columnwidth]{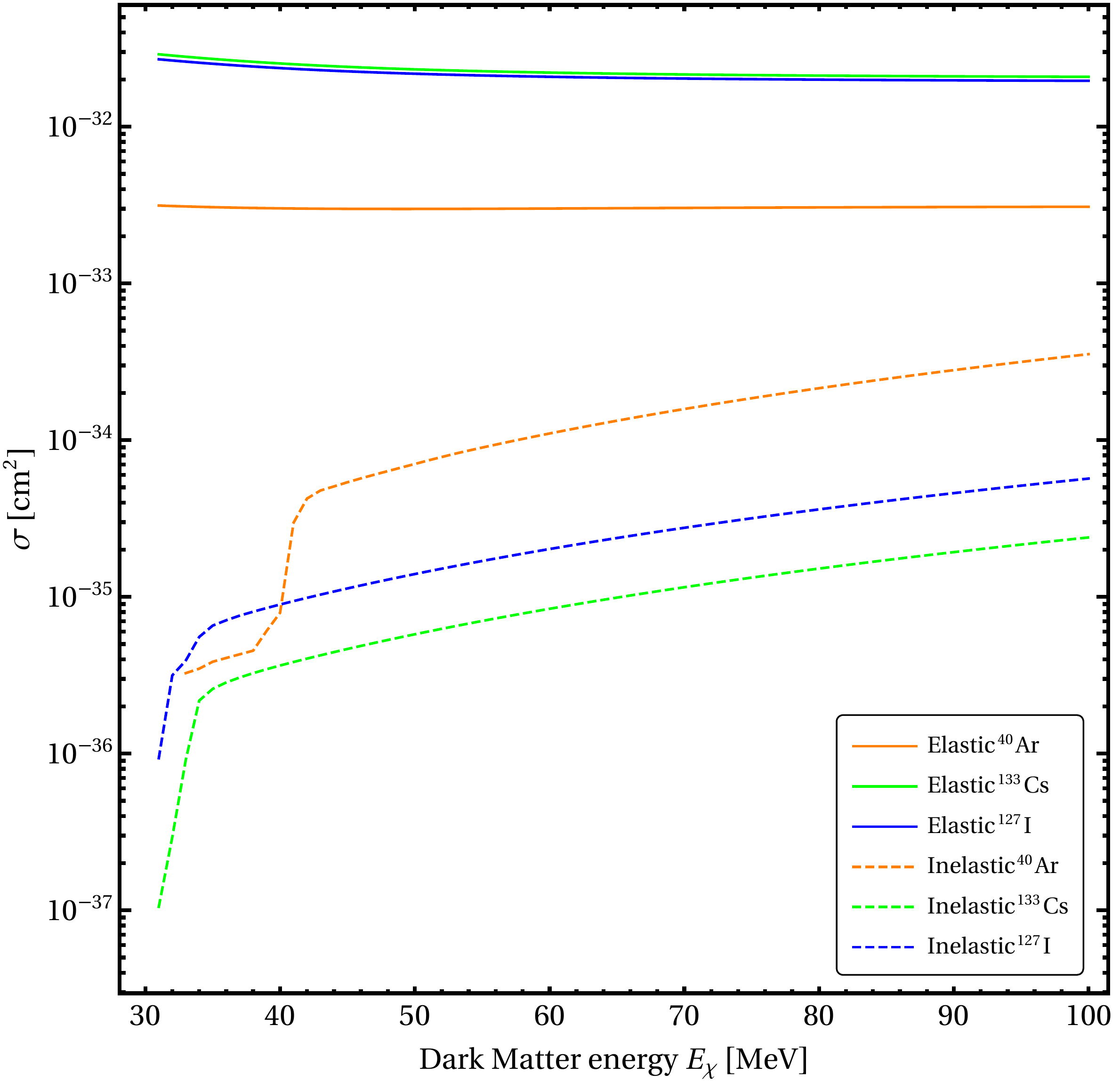}
    \caption{Total elastic and inelastic DM-nucleus scattering cross section for $^{40}$Ar, $^{133}$Cs, and $^{127}$I nuclei with $m_\chi=30$ MeV.}
    \label{fig:DM30}
\end{figure}

In Fig.~\ref{fig:DM30}, we show the elastic and inelastic cross section for dark matter scattering on the target nuclei $^{40}$Ar, $^{133}$Cs and $^{127}$I assuming $m_\chi=30$ MeV, $m_{A'}=90$ MeV, $\epsilon=10^{-4}$  and $g_D=\sqrt{2\pi}$ (this parameter space is allowed by the current experimental data~\cite{COHERENT:2021pvd}). Since the dark matter is not massless, $E_\chi$ has a threshold of 30 MeV, as shown in Fig.~\ref{fig:NuDMspectra} and Fig.~\ref{fig:DM30}. The cross section $\sigma$ is proportional to the couplings constant $\epsilon^2$, so one can scale this plot by changing $\epsilon$. As expected we find that $^{133}$Cs and $^{127}$I have a higher elastic cross section than $^{40}$Ar since $\sigma_{\rm el}$ has explicit $Z^2$ (atomic number) dependence, but have a lower inelastic cross section where $\sigma_{\rm inel}$ is determined by the GT strength, as shown in Fig.~\ref{fig:nuGT30}. The GT transition cross section is the largest for $^{40}$Ar and is the lowest for $^{133}$Cs. There is a plateau for $\chi-^{40}$Ar in Fig.~\ref{fig:DM30}, but not in $^{133}$Cs or $^{127}$I. This is because there are fewer low lying states in $^{40}$Ar than $^{133}$Cs or $^{127}$I. We also observe a small plateau in $\nu-^{40}$Ar in  Fig.~\ref{fig:nuCS}, but this is not as noticeable as with DM scattering since the neutrino is massless. We also calculate the recoil energy spectrum for dark matter scattering in appendix~\ref{app:shellmodel}.

\section{Scattering rates and experimental signatures}

\begin{table*}[th]
    \centering
    \caption{Specifications of the experiments and detectors. Both experiments use a proton beam, and the POT values are expected spills for 5,000 hours of  operation per year.}
    \begin{tabular}{c|c c c| l c c c c}
    \hline \hline
    Experiment     & $E_{\rm beam}$ & POT & Target & Detector: & & &  \\
     & [GeV] & [yr$^{-1}$] & &  target & mass & distance & angle & $E_r^{\rm th}$\\
   \hline
   COHERENT & {1}  & {$1.5\times 10^{23}$} & {Hg} & CsI[Na] & 14.6~kg & 19.3~m & 90$^\circ$ &  6.5~keV \\
   \cite{Akimov:2017ade, COHERENT:2019kwz, coherent2018} & & & & Ar & 24.4~kg & 28.4~m & 137$^\circ$ & 20~keV \\
    CCM~\cite{CCM, CCM:2021leg} & 0.8 & $1.0\times 10^{22}$ & W & Ar & 7~t & 20~m & 90$^\circ$ & 25~keV\\
    \hline \hline
    \end{tabular}
    \label{tab:expspec}
\end{table*}

\begin{figure}[tbh]
    \centering
    \includegraphics[width=0.9\columnwidth]{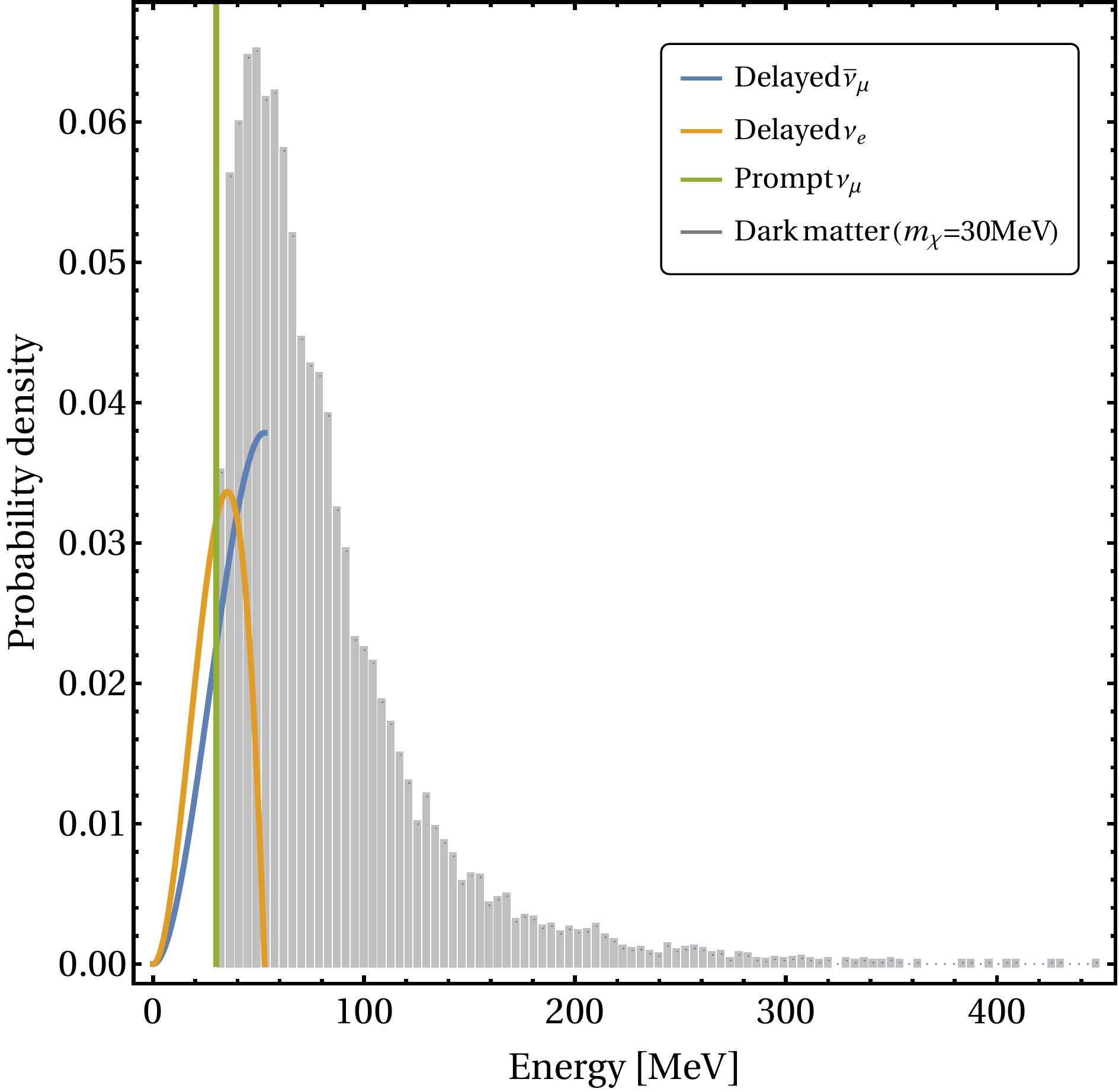}
    \caption{Energy spectra of $\pi$-DAR neutrinos and a sample DM spectrum (in CCM) assuming $m_{A'}=3 m_\chi=90$ MeV.}
    \label{fig:NuDMspectra}
\end{figure}

Using the results of the previous section we compute the rates for neutrino and DM scattering off $^{40}$Ar, $^{133}$Cs, and $^{127}$I nuclei. For a source of neutrinos or DM with flux $\Phi$ [cm$^{-2}$ s$^{-1}$], the number of events expected, $N$, is given by
\begin{align}
    N &= {\mathrm{exposure}} \times \frac{1}{m_T} \times \Phi \int \sigma(E) \nonumber \frac{dP}{dE} dE
\end{align}
where the exposure has dimensions of mass$\times$time, $m_T$ is the target mass, $E$ is the energy of incident particle ($\nu$ or DM) and $\frac{dP}{dE}$ is the corresponding normalized energy distribution (i.e.~$\int \frac{dP}{dE} dE=1$). For neutrinos, we consider pion decay at rest sources, and for DM we simulate the energy spectra in both COHERENT and CCM. For the dark matter spectrum we need to determine the production rates of relevant mesons. We use the results from Ref.~\cite{Dutta:2020vop} which uses GEANT4 to determine these production rates at CCM and  COHERENT. In addition to $\pi^0$ decays and $\pi^-$ absorption, $e^{\pm}$ induced cascades photons are important to produce dark photons for the DM production. We show the DM energy spectrum for CCM (which uses an 800~MeV proton beam), along with the neutrino energy spectra at COHERENT (which uses a 1~GeV proton beam) in Fig.~\ref{fig:NuDMspectra} with $m_\chi=30$ MeV.

Table~\ref{tab:expspec} summarizes the key specifications of the experiments we consider, we additionally assume a detection efficiency of 100\% and that all energy depositions are above threshold. As an example we take the DM mass to be $m_\chi=30$~MeV, dark photon mass $m_{A'}=90$~MeV with coupling constants $\epsilon=10^{-4}$ and $g_D=\sqrt{2\pi}$. We assume that CCM will operate continuously for 3 years with $3\times 10^{22}$ POT, while COHERENT has already ran for $\sim 1.61$ years with their CsI detector ($m_T=14.6$ kg) with $3.20\times 10^{23}$ POT and approximately 0.6 years with their LAr detector ($m_T=24.4$ kg) with $1.34\times 10^{23}$ POT. Table~\ref{tab:events} shows the estimated number of events for the two experiments under these assumptions. The DM energy spectrum has a broader energy range extending up to hundreds of MeV, much higher than the neutrino spectrum. The high energy tail causes DM to induce a higher rate of inelastic events compared to neutrinos. Therefore, the elastic to inelastic event ratios for DM are lower in all detectors.

\begin{table}[H]
\centering
\caption{Number of elastic and NC inelastic neutrino-nucleus, and elastic and inelastic DM-nucleus scattering events ($m_\chi=30$ MeV) for different experimental configurations given in table~\ref{tab:expspec}.}
\label{tab:events}
\begin{tabular}{l|l|l|l|l}
    \hline
    \hline
    Scattering & Experiment & Elastic & Inelastic & Ratio \\
    \hline
    $\nu$-$^{40}$Ar & COHERENT & $2.27\times10^2$  & $3.15$ & $ 7.21 \times 10$ \\
    $\nu$-$^{40}$Ar & CCM & $1.91\times10^4$ & $2.65\times 10^2$ & $7.21 \times 10$ \\
    $\nu$-$^{133}$Cs & COHERENT & $1.16\times 10^3$ & $1.52\times10^{-2} $  & $7.65\times 10^3$ \\
    $\nu$-$^{127}$I & COHERENT & $1.06\times 10^3$ & $3.75\times10^{-1}$ & $2.81\times 10^3$ \\
    \hline
    $\chi$-$^{40}$Ar & COHERENT & $1.18$ & $1.13\times 10^{-1}$ & $1.04\times 10$ \\
    $\chi$-$^{40}$Ar & CCM & $9.92\times10$ & $9.52$ & $1.04\times 10$ \\
    $\chi$-$^{133}$Cs & COHERENT & $4.11$ & $4.91\times 10^{-3}$  & $8.38\times 10^2$ \\
    $\chi$-$^{127}$I & COHERENT & $3.87$ & $1.16\times 10^{-2}$ & $3.33\times 10^2$ \\
    \hline
    \hline
\end{tabular}
\end{table}

\section{Conclusion}
We have applied the nuclear shell model to calculate cross sections of neutrino-nucleus and DM-nucleus scattering. The DM particles are produced from the light vector mediator and this mediator is produced from the kinetic mixing with photons in the stopped-pion experiments (with $\sim 1$ GeV proton beam) we are considering. The number of states we calculated is large enough to include all non-trivial contributions such that the outcome is reliable. In particular we focus on argon, caesium, and iodine nuclei. We computed the cross section in two formalisms: a multipole analysis and the Gamow–Teller operator alone in the long-wavelength limit. We found that Gamow–Teller transitions dominate all other transitions for both neutrino and DM inelastic scattering. Using the Gamow-Teller operator, our calculations show that the inelastic cross section is a few orders of magnitude smaller than the elastic cross section for both neutrino and dark matter scattering. We also computed rates for two experimental setups based on the computed cross section. The inelastic scattering rate is much smaller than the elastic rate, but it will produce a much larger energy deposition that is potentially easier to observe. In the setup we consider, since dark matter spectrum has a higher energy tail than neutrinos, the inelastic contribution is higher and the ratio of the inelastic to the elastic rate can be as large as 0.1 for argon targets. The inelastic neutrino and DM scattering results presented here can be measured in currently running stopped-pion experiments, such as at CCM and COHERENT, and provide additional channels to explore beyond the elastic scattering channel for which they were initially conceived.

\acknowledgments
We thank Calvin W.~Johnson for detailed discussions and light re-writes of his code, Bigstick, which made this work possible. The work of BD and WH are supported in part by the DOE Grant No. DE-SC0010813.  JLN is supported by the Australian Research Council through the ARC Centre of Excellence for Dark Matter Particle Physics, CE200100008. VP acknowledge the support from US DOE under grant DE-SC0009824.

\appendix

\section{DM current derivation} \label{app:dmcurrent}
Some useful kinematics identities
\begin{eqnarray}
\cos\theta &=& {\vec{p_i}^2 +\vec{p_f}^2-2ME_r \over 2\, p_i\, p_f} \\
\vec{p_i} \cdot \vec{p_f} &=&{1\over2}(\vec{p_i}^2 + \vec{p_f}^2-2ME_r)
\end{eqnarray}
where $\vec{p_i}/\vec{p_f}$ is the incoming/outgoing DM momentum, $M$ is the nucleus mass and $E_r$ is the nuclear recoil energy. In the follows we derive the DM currents of Eq.~(\ref{eq:dmInelCS}) in more detail, the DM mass is taken to be $m$:
\begin{align*}
    \sum\limits_{s_i,s_f} l_\mu l_\nu^* &= Tr \left(
    \frac{\cancel{p_f}+m}{2E_f} \, \gamma^\mu \, \frac{\cancel{p_i}+m}{2E_i} \gamma^\nu \right)\\
    &=\frac{1}{4E_f E_i} \left[
    Tr\left(
    \cancel{p_f} \,\gamma^\mu\, \cancel{p_i} \,\gamma^\nu + m^2 \gamma^\mu \gamma^\nu \right)\right]
\end{align*}
where $Tr$ means trace of the matrix. Insert numbers to $\mu$ and $\nu$,
\begin{align*}
    \sum\limits_{s_i,s_f} l_0 l_0^* &= \frac{1}{4E_f E_i} \left[
    Tr\left(
    p_f^\alpha \,\gamma^\alpha \,\gamma^0\, p_i^\beta \,\gamma^\beta \,\gamma^0 + m^2 \gamma^0 \gamma^0 \right)\right] \\
    &=\frac{1}{4E_i E_f} \left[
        4p_f^\alpha p_i^\beta(g^{\alpha 0}g^{\beta 0}-g^{\alpha \beta}g^{00}+g^{\alpha 0}g^{\beta 0}) + m^2 g^{00} \right] \\
    &=\frac{1}{E_i E_f} \left(
        2p_f^0p_i^0 - p_i \cdot p_f + \frac{m^2}{4} \right) \\
    &=\frac{1}{E_i E_f} \left(
        p_f^0p_i^0 + \vec{p_i} \cdot \vec{p_f} + \frac{m^2}{4} \right) \\
    &= 1 + \frac{\vec{p_i}\cdot \vec{p_f}}{E_i E_f} + \frac{m^2}{4E_iE_f} \\
    &= 1 + \frac{\vec{p_i}^2 +\vec{p_f}^2-2ME_r}{2E_i E_f} + \frac{m^2}{4E_i E_f}\\
    &= 1 + \frac{1}{4E_i E_f} \left(2\vec{p_i}^2 +2\vec{p_f}^2-4ME_r+m^2 \right)
\end{align*}

\begin{align*}
    \sum\limits_{s_i,s_f} l_3 l_3^* &= \frac{1}{4E_f E_i} \left[
    Tr\left(
        p_f^\alpha \,\gamma^\alpha \,\gamma^3\, p_i^\beta \,\gamma^\beta \,\gamma^3 + m^2 \gamma^3 \gamma^3 \right)\right] \\
    &=\frac{1}{E_i E_f} \left(
        -2p_f^3p_i^3 + p_i \cdot p_f - \frac{m^2}{4} \right) \\
    &=\frac{1}{E_i E_f} \Big[ -2p_i p_f \cos\theta + \Big( E_iE_f- {1\over2} (\vec{p_i}^2 \\
    &+\vec{p_f}^2-2ME_r) \Big)- {m^2\over4} \Big] \\
    &= 1+ \frac{1}{E_i E_f} \left[ -{3\over2} \left(\vec{p_i}^2 +\vec{p_f}^2 \right) +3ME_r - {m^2\over4} \right] \\
\end{align*}

\begin{align*}
    \sum\limits_{s_i,s_f} l_3 l_0^* &= \frac{1}{4E_f E_i} \left[
    Tr\left(
        p_f^\alpha \,\gamma^\alpha \,\gamma^3\, p_i^\beta \,\gamma^\beta \,\gamma^0 + m^2 \gamma^3 \gamma^0 \right)\right] \\
    &=\frac{p_f^\alpha p_i^\beta}{E_i E_f} \left(
        g^{\alpha 3} g^{\beta 0}-g^{\alpha \beta} g^{30}+g^{\alpha 0}g^{3\beta} \right) \\
    &=-\frac{1}{E_i E_f} \left( p_f^3 p_i^0 + p_f^0 p_i^3 \right) \\
    &=- \left( \frac{p_f^3}{E_f}+\frac{p_i^3}{E_i} \right) \\
    &=-\left({\vec{p_i}^2 +\vec{p_f}^2-2ME_r \over 2p_i E_f}+{p_i\over E_i}\right)
\end{align*}

Summing over $k=1,2,3$
\begin{align*}
    \sum\limits_{s_i,s_f} l_k l_k^* &= \frac{1}{4E_f E_i} \left[
    Tr\left(
        p_f^\alpha \,\gamma^\alpha \,\gamma^k\, p_i^\beta \,\gamma^\beta \,\gamma^k + m^2 \gamma^k \gamma^k \right)\right] \\
    &=\frac{1}{E_i E_f} \left[
        p_f^\alpha p_i^\beta \left( g^{\alpha k}g^{\beta k}-g^{\alpha \beta}g^{kk}+g^{\alpha k}g^{\beta k} \right) + \frac{m^2}{4} g^{kk}\right] \\
    &=\frac{1}{E_i E_f} \left[
    p_f^\alpha p_i^\beta \left( 2g^{\alpha k}g^{\beta k}+3g^{\alpha \beta} \right)-\frac{3}{4}m^2\right] \\
    &=\frac{1}{E_i E_f} \left( 3E_i E_f - \vec{p_i}\cdot \vec{p_f}-\frac{3}{4}m^2 \right)  \\
    &=3- {1\over E_i E_f} \left[ {1\over2} \left(\vec{p_i}^2 +\vec{p_f}^2-2ME_r \right) +{3m^2\over4} \right]
\end{align*}

\begin{align*}
    \sum\limits_{s_i,s_f} &(\vec{l} \times \vec{l}^*)_3 = \sum\limits_{s_i,s_f} l_1 l_2^* - \sum\limits_{s_i,s_f} l_2 l_1^* \\
    &= \frac{1}{4E_i E_f}\left[
    Tr\left(p_f^\alpha \gamma^\alpha \gamma^1 p_i^\beta \gamma^\beta \gamma^2 \right)-Tr\left(p_f^\alpha \gamma^\alpha \gamma^2 p_i^\beta \gamma^\beta \gamma^1 \right)\right]\\
    &=\frac{p_f^\alpha p_i^\beta}{E_i E_f} 
    [ (g^{\alpha 1}g^{\beta 2}-g^{\alpha \beta}g^{12}+g^{\alpha 2}g^{1 \beta}) \\
    &-(g^{\alpha 2}g^{\beta 1}-g^{\alpha \beta}g^{12}+g^{\alpha 1}g^{2 \beta}) ] \\
    &=0
\end{align*}

\section{Shell model results} \label{app:shellmodel}

\begin{figure}[tbh]
    \centering
    \includegraphics[width=0.9\columnwidth]{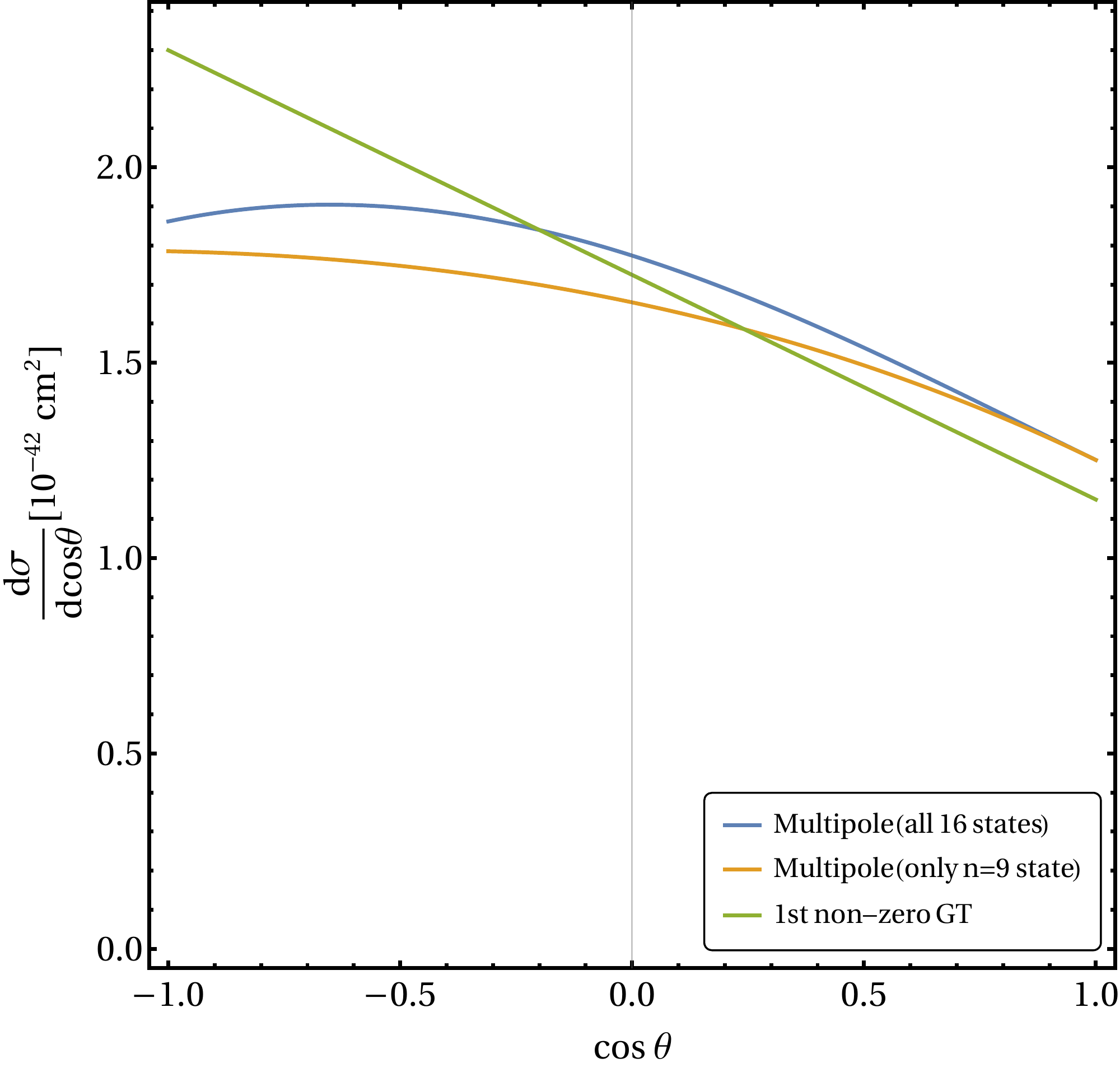}
    \caption{The differential cross section as a function of scattering angle, for NC inelastic $\nu$-$^{40}$Ar scattering with $E_\nu=30$ MeV. The blue and orange curves are calculated using the full cross section formula Eq.~(\ref{xsec}), while the green curve is calculated in the long wavelength limit using Eq.~(\ref{eq:GTnuAngle}).}
    \label{fig:diffxsecangle}
\end{figure}

\begin{figure}[tbh]
    \centering
    \includegraphics[width=0.9\columnwidth]{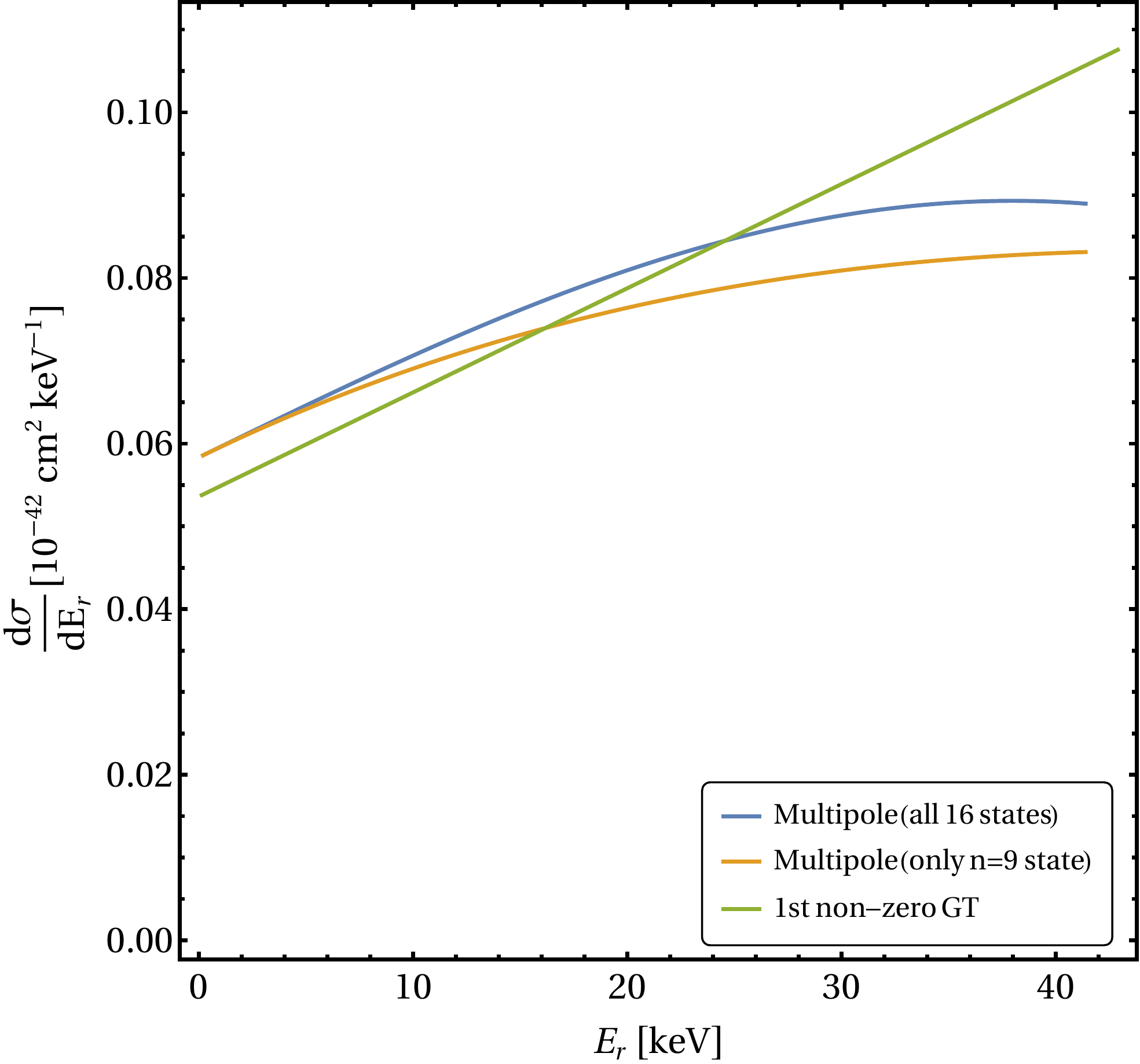}
    \caption{Similar to Fig.~\ref{fig:diffxsecangle}, but differential cross section $\frac{d\sigma}{dE_r}$ plotted as a function of recoil energy $E_r$.}
    \label{fig:diffxsecEr}
\end{figure}

For a more detailed comparison in Fig.~\ref{fig:diffxsecangle} we show the differential neutrino-nucleus cross section as a function of $\cos{\theta}$ for the multipole analysis (including the first 15 transitions) and the GT analysis involving the first transition for a given incoming neutrino energy $E_\nu=30$~MeV. Again we see that the $N=9$ transition, the lowest lying GT transition, dominates the inelastic cross section. Fig.~\ref{fig:diffxsecEr} shows Gamow-Teller cross section in recoil energy $E_r$.

\begin{figure}[h]
    \centering
    \includegraphics[width=0.9\columnwidth]{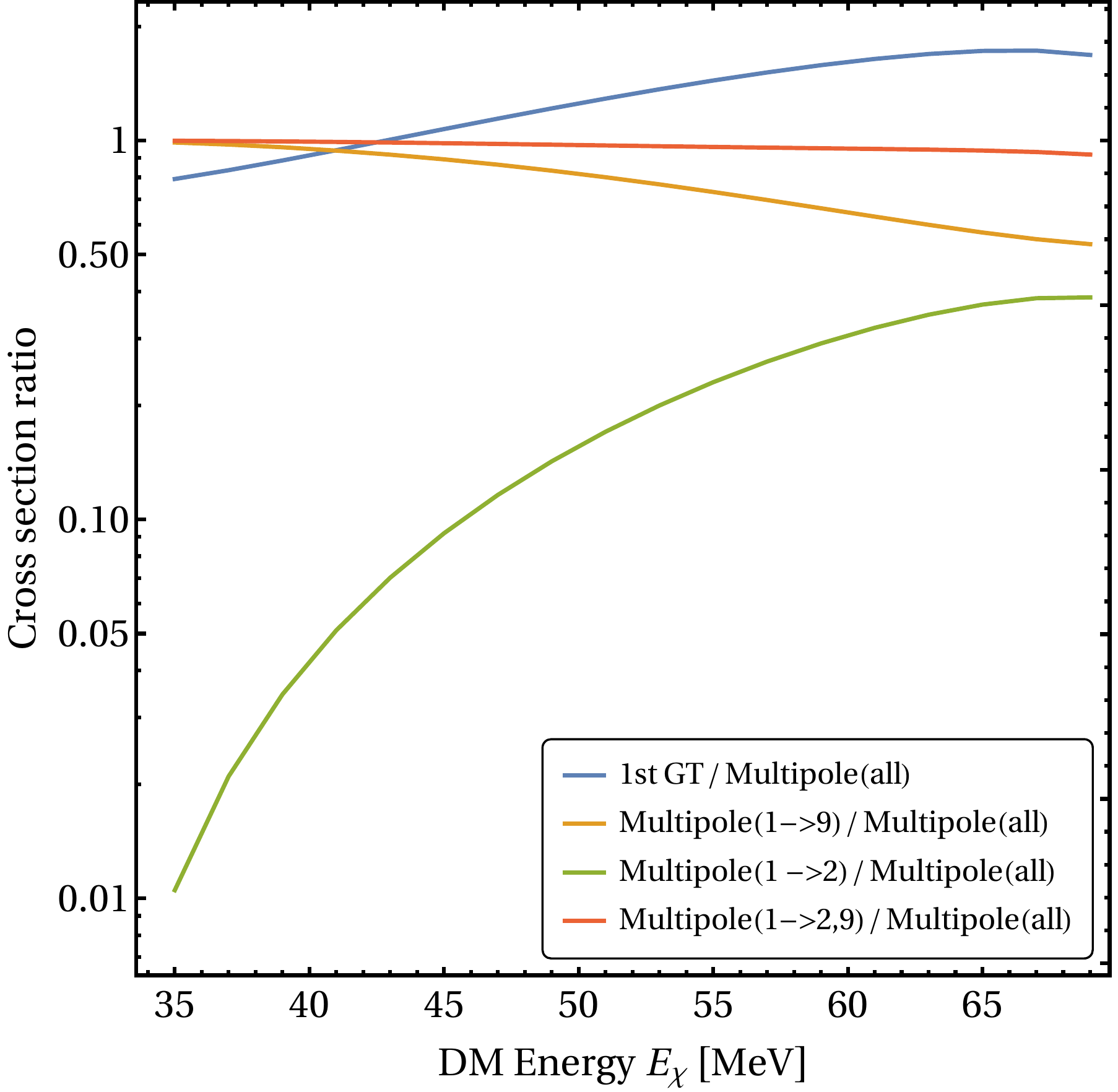}
    \caption{Multipole and GT cross section ratios of $\chi-^{40}$Ar scattering with $m_\chi=30$ MeV.}
    \label{fig:dmGTratio}
\end{figure}

\begin{figure*}[h]
    \centering
    \includegraphics[width=0.9\columnwidth]{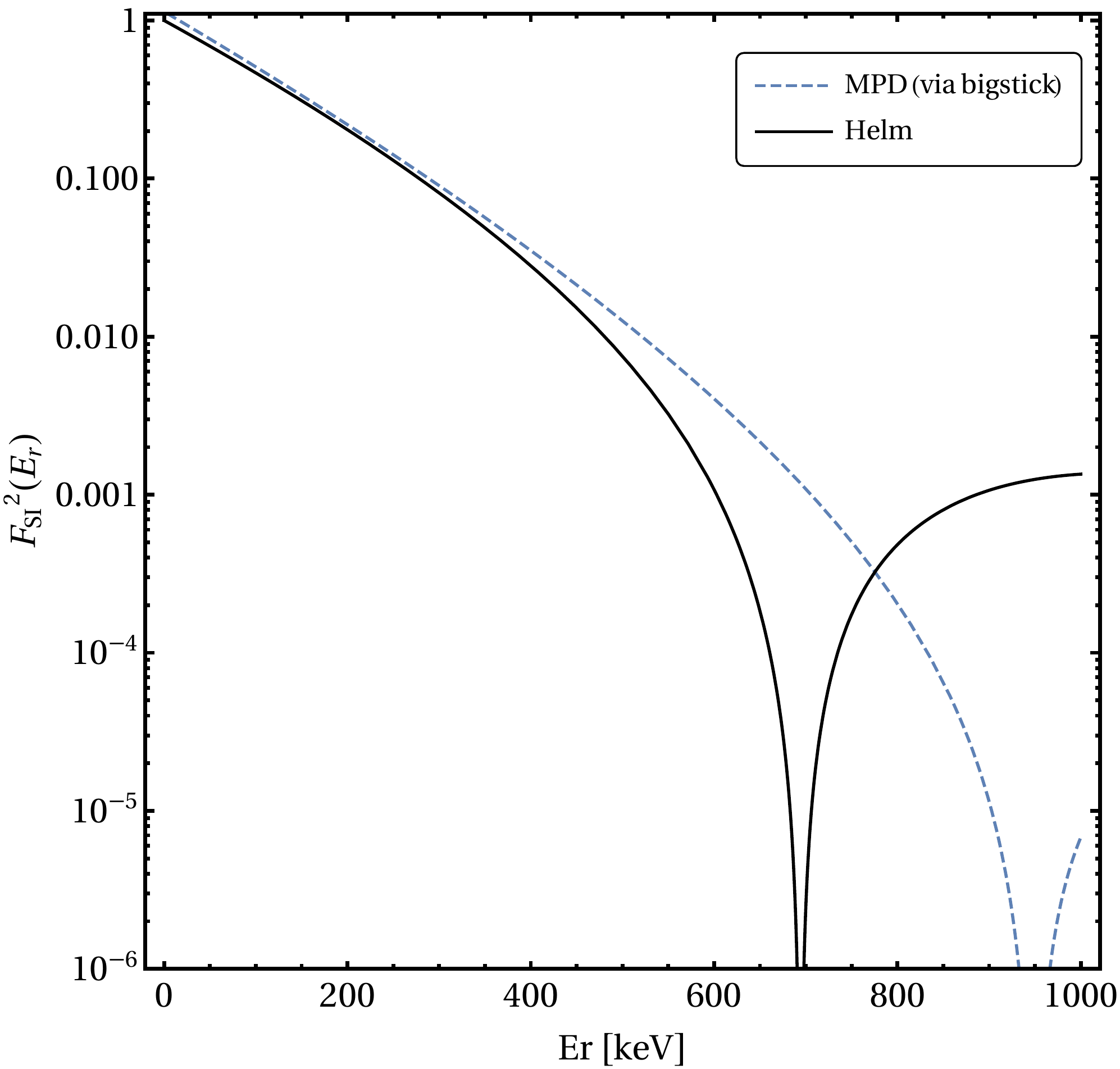} \includegraphics[width=0.9\columnwidth]{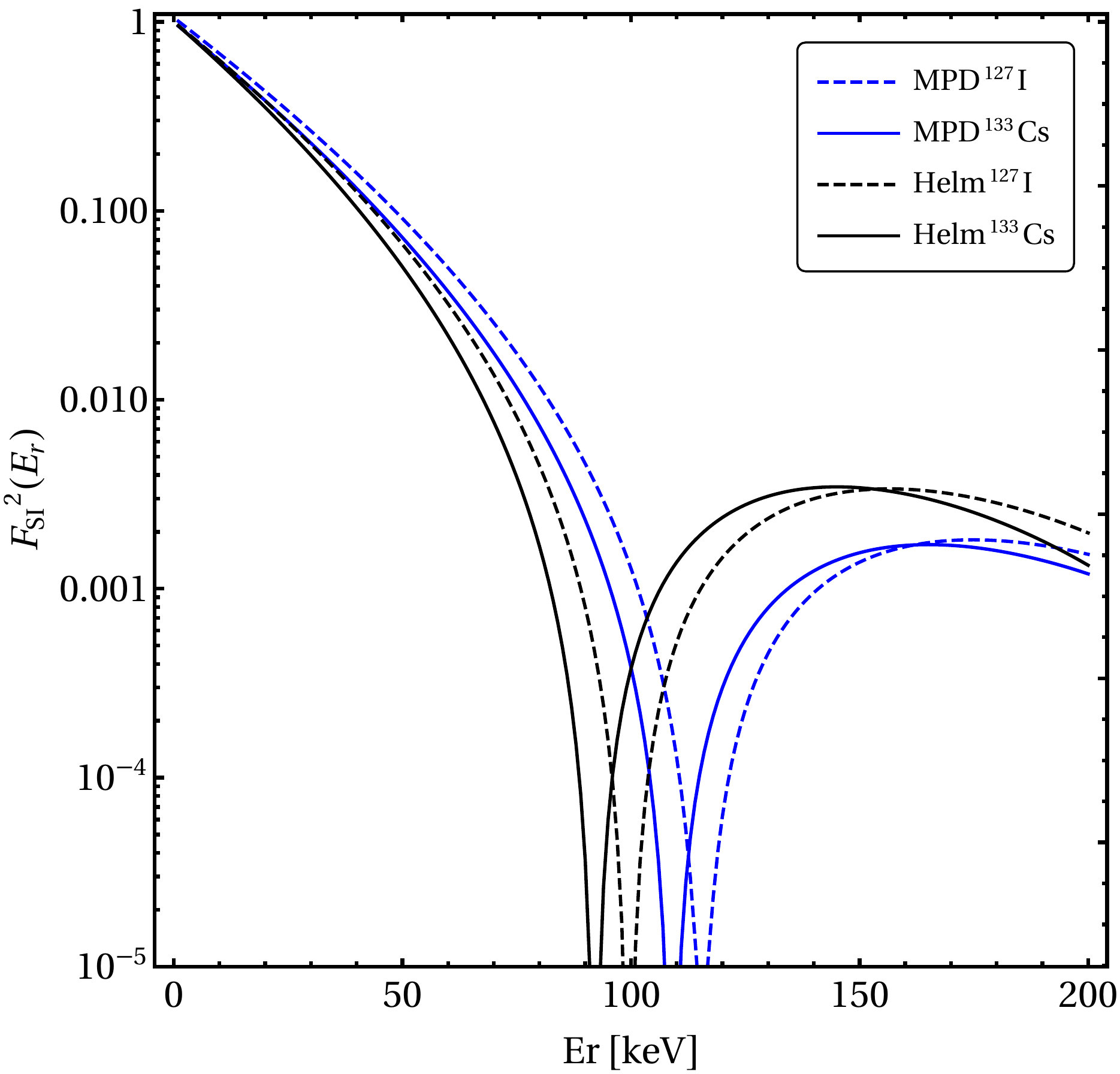} \\
    \caption{Elastic scattering form factors: the Helm parametrization compared with the ground state to ground state transition found via our shell model calculation for $^{40}$Ar (left) and $^{133}$Cs and $^{127}$I (right).}
    \label{fig:ar40gs}
\end{figure*}

Fig.~\ref{fig:dmGTratio} shows multipoles and GT ratios of $\chi-^{40}$Ar scattering cross section. In which Multipole (all) is the sum of all the transitions $1\rightarrow2..16$, dominated by $1\rightarrow2$ and $1\rightarrow9$ transitions. Similar to neutrino case, sum of all the multipole is roughly equivalent to the first GT transition ($1\rightarrow9$). Nevertheless, both $\nu$ and DM match better in low $E_r$/$E_\chi$, while they become less consistent with the long-wave length limit in the relatively higher $E_r$/$E_\chi$ due to the kinematics.

In Fig.~\ref{fig:ar40gs} we plot the shell model {\it{ground state}} to {\it{ground state}} transition compared to the Helm form factor to benchmark the shell-model accuracy and consistency. The difference between our multipole decomposition (MPD) and the Helm form factors is greater than that obtained for the Xe nucleus in~\cite{Vietze:2014vsa} where the GCN5082 interaction was used (note their form factors are plotted as functions of dimensionless $u\equiv q^2 b^2/2$ instead of $E_r$ in Fig.~\ref{fig:ar40gs}). The harmonic oscillator parameter is taken to be $b = \sqrt{41.467/(45A^{-1/3}-25A^{-2/3})}\mathrm{fm}$, which is $\sim2.3$fm for Cs and I. We note that the agreement is good at low momentum transfer, which is most relevant to this work. That said, investigating this discrepancy is a potential line for future work.

\begin{table}[h]
\caption{Nuclear magnetic moments $\mu$ (nm), energy level $E_x$ [keV] and $J^\pi$ spin partiy for the nuclear states of $^{40}$Ar, $^{127}$I, and $^{133}$Cs. Here we label the excited states with $i$, where $i=1$ is the ground state.}
\begin{tabular}{c|c|cc|cc|cc}
\hline
\hline
    Nucleus & $i$ &$J^\pi$ & Expt. & $\mu$ & Expt. & $E_x$ & Expt. \\
    \hline
    $^{127}$I & 1 & $5/2^+$ & $5/2^+$ & 3.851 & 2.813 & 0 & 0 \\
    &2& $7/2^+$ & $7/2^+$ & 3.007 & 2.54 & 37.44 & 57.61 \\
    &3& $3/2^+$ & $3/2^+$ & 0.9155 & 0.97 & 285.9 & 202.86 \\
    & & & & &\\
    $^{133}$Cs& 1 & $7/2^+$ & $7/2^+$ & 3.007 & 2.582 & 0 & 0 \\
    &2&$5/2^+$ &$5/2^+$ & 3.851 & 3.45 & 36.37 & 80.9979 \\
    &3&$5/2^+$ &$5/2^+$ & 2.5849 & 2.0 & 235.36 & 160.6101\\
    & & & & &\\
    $^{40}$Ar& 1 & $0^+$ & $0^+$ & 0 &  & 0 & 0 \\
    &2&$2^+$ &$2^+$ & 0 & -0.04 & 1118.33 & 1460.85 \\
    &3&$2^+$ &$0^+$ & & & 2054.05 & 2120.83 \\
    &4&$4^+$ &$2^+$ & & & 2346.64 & 2524.12 \\
    &5&$0^+$ &$4^+$ & & & 2485.19 & 2892.61 \\
\hline
\hline
\end{tabular}
\label{tab:mag}
\end{table}

\begin{figure}[h]
    \centering
    \includegraphics[width=0.9\columnwidth]{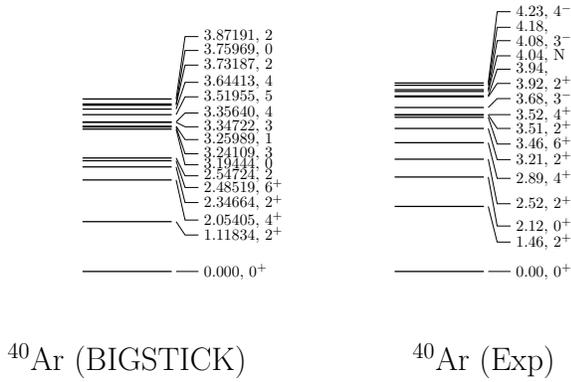}
    \caption{Energy levels and spins of $^{40}$Ar}
    \label{fig:arlvls}
\end{figure}

\begin{figure*}[h]
    \centering
    \includegraphics[width=\columnwidth]{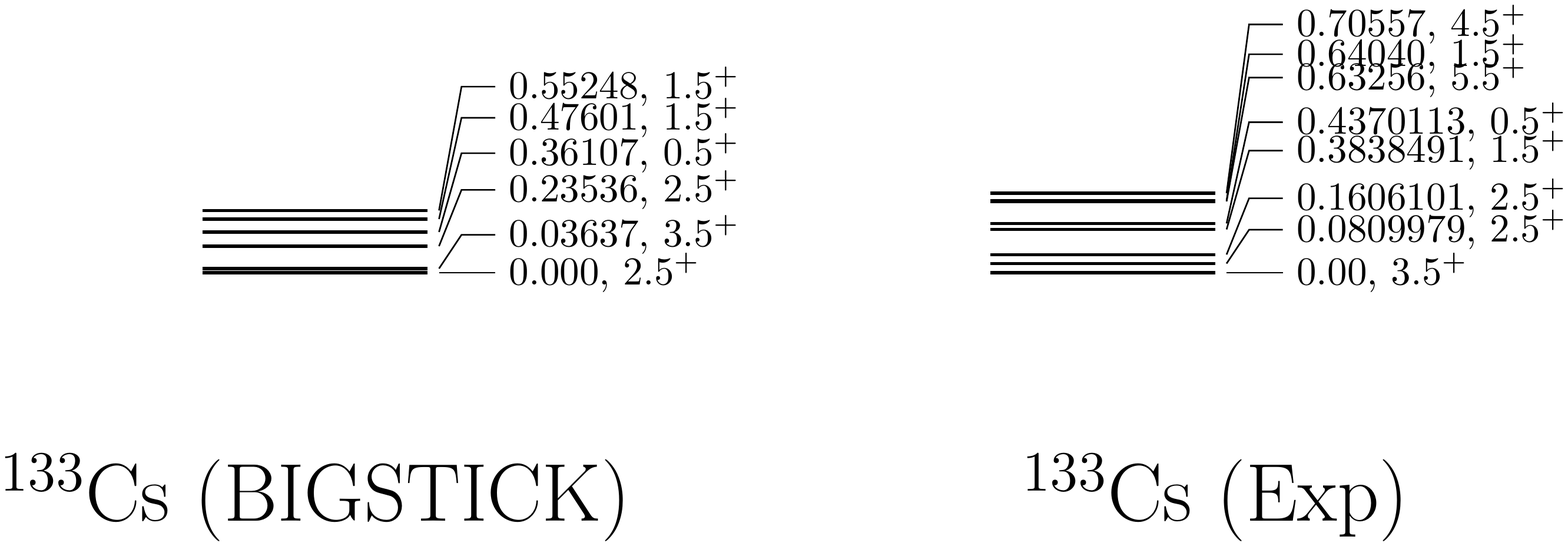}
    \includegraphics[width=\columnwidth]{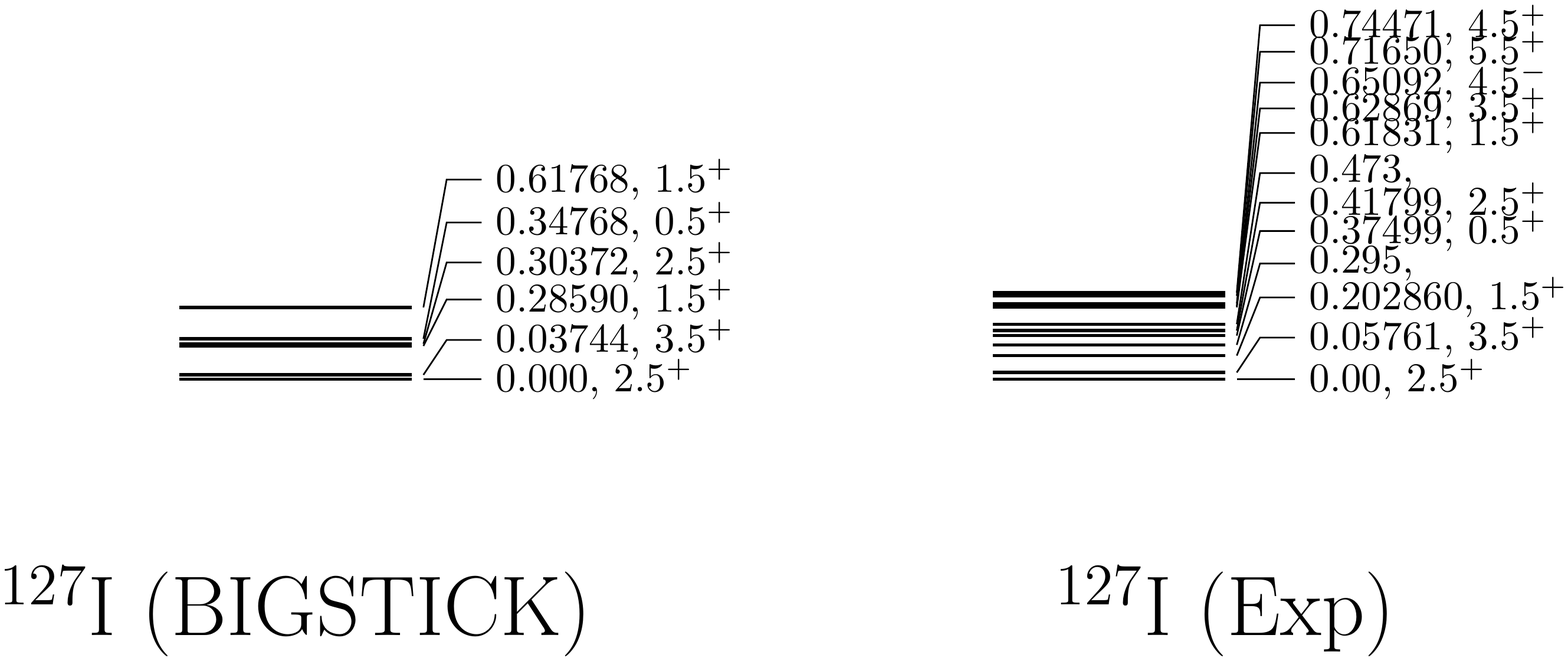}
    \caption{Energy levels and spins of $^{133}$Cs and $^{127}$I}
    \label{fig:csilvls}
\end{figure*}

Table \ref{tab:mag} gives the nuclear magnetic moments $\mu$ (nm) and excitation energies $E_x$ (keV) of the first few states from our shell model calculation, which compare reasonably well with the experimental values~\cite{Sahu:2020kwh,NDS}. Our shell model predicts that the first excited state of $^{40}$Ar has 2 protons moved from 0d$_{3/2}$ (ground state) to 0f$_{7/2}$, so every nucleon is paired, which implies $\mu=0$.

\bibliographystyle{bibi}
\bibliography{civns_references}

\providecommand{\href}[2]{#2}\begingroup\raggedright\begin{thebibliography}{10}

\bibitem{Akimov:2017ade}
{\scshape COHERENT} Collaboration, D.~Akimov et~al., \emph{{Observation of
  Coherent Elastic Neutrino-Nucleus Scattering}},
  \href{https://doi.org/10.1126/science.aao0990}{\emph{Science} {\bfseries 357}
  (2017) 1123} [\href{https://arxiv.org/abs/1708.01294}{{\ttfamily
  1708.01294}}].

\bibitem{CCM}
R.~{van de Water} and {Coherent-Mills Experiment Team}, \emph{{Searching for
  Sterile Neutrinos with the Coherent CAPTAIN-Mills Detector at the Los Alamos
  Neutron Science Center}},  in \emph{APS April Meeting Abstracts}, vol.~2019
  of \emph{APS Meeting Abstracts}, p.~Z14.009, Jan., 2019.

\bibitem{Akimov:2021dab}
D.~Akimov et~al., \emph{{Measurement of the Coherent Elastic Neutrino-Nucleus
  Scattering Cross Section on CsI by COHERENT}},
  \href{https://arxiv.org/abs/2110.07730}{{\ttfamily 2110.07730}}.

\bibitem{Ohlsson:2012kf}
T.~Ohlsson, \emph{{Status of non-standard neutrino interactions}},
  \href{https://doi.org/10.1088/0034-4885/76/4/044201}{\emph{Rept. Prog. Phys.}
  {\bfseries 76} (2013) 044201}
  [\href{https://arxiv.org/abs/1209.2710}{{\ttfamily 1209.2710}}].

\bibitem{Miranda:2015dra}
O.~G. Miranda and H.~Nunokawa, \emph{{Non standard neutrino interactions:
  current status and future prospects}},
  \href{https://doi.org/10.1088/1367-2630/17/9/095002}{\emph{New J. Phys.}
  {\bfseries 17} (2015) 095002}
  [\href{https://arxiv.org/abs/1505.06254}{{\ttfamily 1505.06254}}].

\bibitem{Dent:2017mpr}
J.~B. Dent, B.~Dutta, S.~Liao, J.~L. Newstead, L.~E. Strigari and J.~W. Walker,
  \emph{{Accelerator and reactor complementarity in coherent neutrino-nucleus
  scattering}}, \href{https://doi.org/10.1103/PhysRevD.97.035009}{\emph{Phys.
  Rev. D} {\bfseries 97} (2018) 035009}
  [\href{https://arxiv.org/abs/1711.03521}{{\ttfamily 1711.03521}}].

\bibitem{Liao:2017uzy}
J.~Liao and D.~Marfatia, \emph{{COHERENT constraints on nonstandard neutrino
  interactions}},
  \href{https://doi.org/10.1016/j.physletb.2017.10.046}{\emph{Phys. Lett. B}
  {\bfseries 775} (2017) 54}
  [\href{https://arxiv.org/abs/1708.04255}{{\ttfamily 1708.04255}}].

\bibitem{Denton:2018xmq}
P.~B. Denton, Y.~Farzan and I.~M. Shoemaker, \emph{{Testing large non-standard
  neutrino interactions with arbitrary mediator mass after COHERENT data}},
  \href{https://doi.org/10.1007/JHEP07(2018)037}{\emph{JHEP} {\bfseries 07}
  (2018) 037} [\href{https://arxiv.org/abs/1804.03660}{{\ttfamily
  1804.03660}}].

\bibitem{Billard:2018jnl}
J.~Billard, J.~Johnston and B.~J. Kavanagh, \emph{{Prospects for exploring New
  Physics in Coherent Elastic Neutrino-Nucleus Scattering}},
  \href{https://doi.org/10.1088/1475-7516/2018/11/016}{\emph{JCAP} {\bfseries
  11} (2018) 016} [\href{https://arxiv.org/abs/1805.01798}{{\ttfamily
  1805.01798}}].

\bibitem{Altmannshofer:2018xyo}
W.~Altmannshofer, M.~Tammaro and J.~Zupan, \emph{{Non-standard neutrino
  interactions and low energy experiments}},
  \href{https://doi.org/10.1007/JHEP11(2021)113}{\emph{JHEP} {\bfseries 09}
  (2019) 083} [\href{https://arxiv.org/abs/1812.02778}{{\ttfamily
  1812.02778}}]. [Erratum: JHEP 11, 113 (2021)].

\bibitem{Dutta:2019eml}
B.~Dutta, S.~Liao, S.~Sinha and L.~E. Strigari, \emph{{Searching for Beyond the
  Standard Model Physics with COHERENT Energy and Timing Data}},
  \href{https://doi.org/10.1103/PhysRevLett.123.061801}{\emph{Phys. Rev. Lett.}
  {\bfseries 123} (2019) 061801}
  [\href{https://arxiv.org/abs/1903.10666}{{\ttfamily 1903.10666}}].

\bibitem{Canas:2019fjw}
B.~C. Canas, E.~A. Garces, O.~G. Miranda, A.~Parada and G.~Sanchez~Garcia,
  \emph{{Interplay between nonstandard and nuclear constraints in coherent
  elastic neutrino-nucleus scattering experiments}},
  \href{https://doi.org/10.1103/PhysRevD.101.035012}{\emph{Phys. Rev. D}
  {\bfseries 101} (2020) 035012}
  [\href{https://arxiv.org/abs/1911.09831}{{\ttfamily 1911.09831}}].

\bibitem{Khan:2019cvi}
A.~N. Khan and W.~Rodejohann, \emph{{New physics from COHERENT data with an
  improved quenching factor}},
  \href{https://doi.org/10.1103/PhysRevD.100.113003}{\emph{Phys. Rev. D}
  {\bfseries 100} (2019) 113003}
  [\href{https://arxiv.org/abs/1907.12444}{{\ttfamily 1907.12444}}].

\bibitem{Coloma:2020nhf}
P.~Coloma, I.~Esteban, M.~C. Gonzalez-Garcia and J.~Menendez,
  \emph{{Determining the nuclear neutron distribution from Coherent Elastic
  neutrino-Nucleus Scattering: current results and future prospects}},
  \href{https://doi.org/10.1007/JHEP08(2020)030}{\emph{JHEP} {\bfseries 08}
  (2020) 030} [\href{https://arxiv.org/abs/2006.08624}{{\ttfamily
  2006.08624}}].

\bibitem{Denton:2020hop}
P.~B. Denton and J.~Gehrlein, \emph{{A Statistical Analysis of the COHERENT
  Data and Applications to New Physics}},
  \href{https://doi.org/10.1007/JHEP04(2021)266}{\emph{JHEP} {\bfseries 04}
  (2021) 266} [\href{https://arxiv.org/abs/2008.06062}{{\ttfamily
  2008.06062}}].

\bibitem{Dutta:2020che}
B.~Dutta, R.~F. Lang, S.~Liao, S.~Sinha, L.~Strigari and A.~Thompson, \emph{{A
  global analysis strategy to resolve neutrino NSI degeneracies with scattering
  and oscillation data}},
  \href{https://doi.org/10.1007/JHEP09(2020)106}{\emph{JHEP} {\bfseries 09}
  (2020) 106} [\href{https://arxiv.org/abs/2002.03066}{{\ttfamily
  2002.03066}}].

\bibitem{deNiverville:2011it}
P.~deNiverville, M.~Pospelov and A.~Ritz, \emph{{Observing a light dark matter
  beam with neutrino experiments}},
  \href{https://doi.org/10.1103/PhysRevD.84.075020}{\emph{Phys. Rev. D}
  {\bfseries 84} (2011) 075020}
  [\href{https://arxiv.org/abs/1107.4580}{{\ttfamily 1107.4580}}].

\bibitem{deNiverville:2015mwa}
P.~deNiverville, M.~Pospelov and A.~Ritz, \emph{{Light new physics in coherent
  neutrino-nucleus scattering experiments}},
  \href{https://doi.org/10.1103/PhysRevD.92.095005}{\emph{Phys. Rev.}
  {\bfseries D92} (2015) 095005}
  [\href{https://arxiv.org/abs/1505.07805}{{\ttfamily 1505.07805}}].

\bibitem{Ge:2017mcq}
S.-F. Ge and I.~M. Shoemaker, \emph{{Constraining Photon Portal Dark Matter
  with Texono and Coherent Data}},
  \href{https://doi.org/10.1007/JHEP11(2018)066}{\emph{JHEP} {\bfseries 11}
  (2018) 066} [\href{https://arxiv.org/abs/1710.10889}{{\ttfamily
  1710.10889}}].

\bibitem{Dutta:PRL2020}
B.~Dutta, D.~Kim, S.~Liao, J.-C. Park, S.~Shin and L.~E. Strigari, \emph{Dark
  matter signals from timing spectra at neutrino experiments},
  \href{https://doi.org/10.1103/PhysRevLett.124.121802}{\emph{Phys. Rev. Lett.}
  {\bfseries 124} (2020) 121802}.

\bibitem{COHERENT:2021pvd}
{\scshape COHERENT} Collaboration, D.~Akimov et~al., \emph{{First Probe of
  Sub-GeV Dark Matter Beyond the Cosmological Expectation with the COHERENT CsI
  Detector at the SNS}},  \href{https://arxiv.org/abs/2110.11453}{{\ttfamily
  2110.11453}}.

\bibitem{COHERENT:2019kwz}
{\scshape COHERENT} Collaboration, D.~Akimov et~al., \emph{{Sensitivity of the
  COHERENT Experiment to Accelerator-Produced Dark Matter}},
  \href{https://doi.org/10.1103/PhysRevD.102.052007}{\emph{Phys. Rev. D}
  {\bfseries 102} (2020) 052007}
  [\href{https://arxiv.org/abs/1911.06422}{{\ttfamily 1911.06422}}].

\bibitem{Aguilar-Arevalo:2021sbh}
A.~A. Aguilar-Arevalo et~al., \emph{{First Leptophobic Dark Matter Search from
  Coherent CAPTAIN-Mills}},  \href{https://arxiv.org/abs/2109.14146}{{\ttfamily
  2109.14146}}.

\bibitem{CCM:2021leg}
{\scshape CCM} Collaboration, A.~A. Aguilar-Arevalo et~al., \emph{{First Dark
  Matter Search Results From Coherent CAPTAIN-Mills}},
  \href{https://arxiv.org/abs/2105.14020}{{\ttfamily 2105.14020}}.

\bibitem{DUNE:2020zfm}
{\scshape DUNE} Collaboration, B.~Abi et~al., \emph{{Supernova neutrino burst
  detection with the Deep Underground Neutrino Experiment}},
  \href{https://doi.org/10.1140/epjc/s10052-021-09166-w}{\emph{Eur. Phys. J. C}
  {\bfseries 81} (2021) 423}
  [\href{https://arxiv.org/abs/2008.06647}{{\ttfamily 2008.06647}}].

\bibitem{Hyper-Kamiokande:2016srs}
{\scshape Hyper-Kamiokande} Collaboration, K.~Abe et~al., \emph{{Physics
  potentials with the second Hyper-Kamiokande detector in Korea}},
  \href{https://doi.org/10.1093/ptep/pty044}{\emph{PTEP} {\bfseries 2018}
  (2018) 063C01} [\href{https://arxiv.org/abs/1611.06118}{{\ttfamily
  1611.06118}}].

\bibitem{Maschuw:1998qh}
{\scshape KARMEN} Collaboration, R.~Maschuw, \emph{{Neutrino spectroscopy with
  KARMEN}}, \href{https://doi.org/10.1016/S0146-6410(98)00024-6}{\emph{Prog.
  Part. Nucl. Phys.} {\bfseries 40} (1998) 183}.

\bibitem{LSND:2001fbw}
{\scshape LSND} Collaboration, L.~B. Auerbach et~al., \emph{{Measurements of
  charged current reactions of nu(e) on 12-C}},
  \href{https://doi.org/10.1103/PhysRevC.64.065501}{\emph{Phys. Rev. C}
  {\bfseries 64} (2001) 065501}
  [\href{https://arxiv.org/abs/hep-ex/0105068}{{\ttfamily hep-ex/0105068}}].

\bibitem{Haxton:1987kc}
W.~C. Haxton, \emph{{The Nuclear Response of Water Cherenkov Detectors to
  Supernova and Solar Neutrinos}},
  \href{https://doi.org/10.1103/PhysRevD.36.2283}{\emph{Phys. Rev. D}
  {\bfseries 36} (1987) 2283}.

\bibitem{Kolbe:1992xu}
E.~Kolbe, K.~Langanke, S.~Krewald and F.~K. Thielemann, \emph{{Inelastic
  neutrino scattering on C-12 and O-16 above the particle emission threshold}},
  \href{https://doi.org/10.1016/0375-9474(92)90175-J}{\emph{Nucl. Phys. A}
  {\bfseries 540} (1992) 599}.

\bibitem{Ormand:1994js}
W.~E. Ormand, P.~M. Pizzochero, P.~F. Bortignon and R.~A. Broglia,
  \emph{{Neutrino capture cross-sections for Ar-40 and Beta decay of Ti-40}},
  \href{https://doi.org/10.1016/0370-2693(94)01605-C}{\emph{Phys. Lett. B}
  {\bfseries 345} (1995) 343}
  [\href{https://arxiv.org/abs/nucl-th/9405007}{{\ttfamily nucl-th/9405007}}].

\bibitem{Kolbe:1999vc}
E.~Kolbe, K.~Langanke and G.~Martinez-Pinedo, \emph{{The Inclusive Fe-56
  (electron neutrino, e-) Co-56 cross-section}},
  \href{https://doi.org/10.1103/PhysRevC.60.052801}{\emph{Phys. Rev. C}
  {\bfseries 60} (1999) 052801}
  [\href{https://arxiv.org/abs/nucl-th/9905001}{{\ttfamily nucl-th/9905001}}].

\bibitem{Hayes:1999ew}
A.~C. Hayes and I.~S. Towner, \emph{{Shell model calculations of neutrino
  scattering from C-12}},
  \href{https://doi.org/10.1103/PhysRevC.61.044603}{\emph{Phys. Rev. C}
  {\bfseries 61} (2000) 044603}
  [\href{https://arxiv.org/abs/nucl-th/9907049}{{\ttfamily nucl-th/9907049}}].

\bibitem{Volpe:2000zn}
C.~Volpe, N.~Auerbach, G.~Colo, T.~Suzuki and N.~Van~Giai, \emph{{Microscopic
  theories of neutrino C-12 reactions}},
  \href{https://doi.org/10.1103/PhysRevC.62.015501}{\emph{Phys. Rev. C}
  {\bfseries 62} (2000) 015501}
  [\href{https://arxiv.org/abs/nucl-th/0001050}{{\ttfamily nucl-th/0001050}}].

\bibitem{Engel:2002hg}
J.~Engel, G.~C. McLaughlin and C.~Volpe, \emph{{What can be learned with a lead
  based supernova neutrino detector?}},
  \href{https://doi.org/10.1103/PhysRevD.67.013005}{\emph{Phys. Rev. D}
  {\bfseries 67} (2003) 013005}
  [\href{https://arxiv.org/abs/hep-ph/0209267}{{\ttfamily hep-ph/0209267}}].

\bibitem{Suzuki:2006qd}
T.~Suzuki, S.~Chiba, T.~Yoshida, T.~Kajino and T.~Otsuka, \emph{{Neutrino
  nucleus reactions based on new shell model Hamiltonians}},
  \href{https://doi.org/10.1103/PhysRevC.74.034307}{\emph{Phys. Rev. C}
  {\bfseries 74} (2006) 034307}
  [\href{https://arxiv.org/abs/nucl-th/0608056}{{\ttfamily nucl-th/0608056}}].

\bibitem{Suzuki:2009zzc}
T.~Suzuki, M.~Honma, K.~Higashiyama, T.~Yoshida, T.~Kajino, T.~Otsuka, H.~Umeda
  and K.~Nomoto, \emph{{Neutrino-induced reactions on Fe-56 and Ni-56, and
  production of Mn-55 in population III stars}},
  \href{https://doi.org/10.1103/PhysRevC.79.061603}{\emph{Phys. Rev. C}
  {\bfseries 79} (2009) 061603}.

\bibitem{Cheoun:2011hj}
M.-K. Cheoun, E.~Ha, T.~Hayakawa, S.~Chiba, K.~Nakamura, T.~Kajino and G.~J.
  Mathews, \emph{{Neutrino induced reactions related to the $\nu$-process
  nucleosynthesis of ${}^{92}$Nb and ${}^{98}$Tc}},
  \href{https://doi.org/10.1103/PhysRevC.85.065807}{\emph{Phys. Rev. C}
  {\bfseries 85} (2012) 065807}
  [\href{https://arxiv.org/abs/1108.4229}{{\ttfamily 1108.4229}}].

\bibitem{Kostensalo:2018kgh}
J.~Kostensalo, J.~Suhonen and K.~Zuber, \emph{{Shell-model computed cross
  sections for charged-current scattering of astrophysical neutrinos off
  $^{40}$Ar}}, \href{https://doi.org/10.1103/PhysRevC.97.034309}{\emph{Phys.
  Rev. C} {\bfseries 97} (2018) 034309}.

\bibitem{VanDessel:2019atx}
N.~Van~Dessel, N.~Jachowicz and A.~Nikolakopoulos, \emph{{Forbidden transitions
  in neutral and charged current interactions between low-energy neutrinos and
  Argon}}, \href{https://doi.org/10.1103/PhysRevC.100.055503}{\emph{Phys. Rev.
  C} {\bfseries 100} (2019) 055503}
  [\href{https://arxiv.org/abs/1903.07726}{{\ttfamily 1903.07726}}].

\bibitem{VanDessel:2020epd}
N.~Van~Dessel, V.~Pandey, H.~Ray and N.~Jachowicz, \emph{{Nuclear Structure
  Physics in Coherent Elastic Neutrino-Nucleus Scattering}},
  \href{https://arxiv.org/abs/2007.03658}{{\ttfamily 2007.03658}}.

\bibitem{Sahu:2020kwh}
R.~Sahu, D.~K. Papoulias, V.~K.~B. Kota and T.~S. Kosmas, \emph{{Elastic and
  inelastic scattering of neutrinos and weakly interacting massive particles on
  nuclei}}, \href{https://doi.org/10.1103/PhysRevC.102.035501}{\emph{Phys. Rev.
  C} {\bfseries 102} (2020) 035501}
  [\href{https://arxiv.org/abs/2004.04055}{{\ttfamily 2004.04055}}].

\bibitem{Bednyakov:2018PRD}
V.~A. Bednyakov and D.~V. Naumov, \emph{Coherency and incoherency in
  neutrino-nucleus elastic and inelastic scattering},
  \href{https://doi.org/10.1103/PhysRevD.98.053004}{\emph{Phys. Rev. D}
  {\bfseries 98} (2018) 053004}.

\bibitem{Nikolakopoulos:2020alk}
A.~Nikolakopoulos, V.~Pandey, J.~Spitz and N.~Jachowicz, \emph{{Modeling
  quasielastic interactions of monoenergetic kaon decay-at-rest neutrinos}},
  \href{https://doi.org/10.1103/PhysRevC.103.064603}{\emph{Phys. Rev. C}
  {\bfseries 103} (2021) 064603}
  [\href{https://arxiv.org/abs/2010.05794}{{\ttfamily 2010.05794}}].

\bibitem{Pandey:2014tza}
V.~Pandey, N.~Jachowicz, T.~Van~Cuyck, J.~Ryckebusch and M.~Martini,
  \emph{{Low-energy excitations and quasielastic contribution to
  electron-nucleus and neutrino-nucleus scattering in the continuum
  random-phase approximation}},
  \href{https://doi.org/10.1103/PhysRevC.92.024606}{\emph{Phys. Rev. C}
  {\bfseries 92} (2015) 024606}
  [\href{https://arxiv.org/abs/1412.4624}{{\ttfamily 1412.4624}}].

\bibitem{Dutta:2020vop}
B.~Dutta, D.~Kim, S.~Liao, J.-C. Park, S.~Shin, L.~E. Strigari and A.~Thompson,
  \emph{{Searching for dark matter signals in timing spectra at neutrino
  experiments}}, \href{https://doi.org/10.1007/JHEP01(2022)144}{\emph{JHEP}
  {\bfseries 01} (2022) 144}
  [\href{https://arxiv.org/abs/2006.09386}{{\ttfamily 2006.09386}}].

\bibitem{Dutta:2021cip}
B.~Dutta, D.~Kim, A.~Thompson, R.~T. Thornton and R.~G. Van~de Water,
  \emph{{Solutions to the MiniBooNE Anomaly from New Physics in Charged Meson
  Decays}},  \href{https://arxiv.org/abs/2110.11944}{{\ttfamily 2110.11944}}.

\bibitem{Baudis:2013bba}
L.~Baudis, G.~Kessler, P.~Klos, R.~F. Lang, J.~Men\'endez, S.~Reichard and
  A.~Schwenk, \emph{{Signatures of Dark Matter Scattering Inelastically Off
  Nuclei}}, \href{https://doi.org/10.1103/PhysRevD.88.115014}{\emph{Phys. Rev.
  D} {\bfseries 88} (2013) 115014}
  [\href{https://arxiv.org/abs/1309.0825}{{\ttfamily 1309.0825}}].

\bibitem{Vietze:2014vsa}
L.~Vietze, P.~Klos, J.~Men\'endez, W.~C. Haxton and A.~Schwenk, \emph{{Nuclear
  structure aspects of spin-independent WIMP scattering off xenon}},
  \href{https://doi.org/10.1103/PhysRevD.91.043520}{\emph{Phys. Rev. D}
  {\bfseries 91} (2015) 043520}
  [\href{https://arxiv.org/abs/1412.6091}{{\ttfamily 1412.6091}}].

\bibitem{Helm:1956zz}
R.~H. Helm, \emph{{Inelastic and Elastic Scattering of 187-Mev Electrons from
  Selected Even-Even Nuclei}},
  \href{https://doi.org/10.1103/PhysRev.104.1466}{\emph{Phys. Rev.} {\bfseries
  104} (1956) 1466}.

\bibitem{Hoferichter:2020osn}
M.~Hoferichter, J.~Men\'endez and A.~Schwenk, \emph{{Coherent elastic
  neutrino-nucleus scattering: EFT analysis and nuclear responses}},
  \href{https://doi.org/10.1103/PhysRevD.102.074018}{\emph{Phys. Rev. D}
  {\bfseries 102} (2020) 074018}
  [\href{https://arxiv.org/abs/2007.08529}{{\ttfamily 2007.08529}}].

\bibitem{Tomalak:2020zfh}
O.~Tomalak, P.~Machado, V.~Pandey and R.~Plestid, \emph{{Flavor-dependent
  radiative corrections in coherent elastic neutrino-nucleus scattering}},
  \href{https://doi.org/10.1007/JHEP02(2021)097}{\emph{JHEP} {\bfseries 02}
  (2021) 097} [\href{https://arxiv.org/abs/2011.05960}{{\ttfamily
  2011.05960}}].

\bibitem{DeForest:1966ycn}
T.~De~Forest, Jr. and J.~D. Walecka, \emph{{Electron scattering and nuclear
  structure}}, \href{https://doi.org/10.1080/00018736600101254}{\emph{Adv.
  Phys.} {\bfseries 15} (1966) 1}.

\bibitem{Serot:1978vj}
B.~D. Serot, \emph{{Semileptonic Weak and Electromagnetic Interactions with
  Nuclei: Nuclear Current Operators Through Order (v/c)**2 (Nucleon)}},
  \href{https://doi.org/10.1016/0375-9474(78)90561-4}{\emph{Nucl. Phys. A}
  {\bfseries 308} (1978) 457}.

\bibitem{Donnelly:1979ezn}
T.~W. Donnelly and W.~C. Haxton, \emph{{Multipole operators in semileptonic
  weak and electromagnetic interactions with nuclei}},
  \href{https://doi.org/10.1016/0092-640X(79)90003-2}{\emph{Atom. Data Nucl.
  Data Tabl.} {\bfseries 23} (1979) 103}.

\bibitem{walecka2004theoretical}
J.~Walecka, \emph{Theoretical Nuclear and Subnuclear Physics}, Theoretical
  Nuclear and Subnuclear Physics. Imperial College Press, 2004.

\bibitem{Haxton:2008zza}
W.~Haxton and C.~Lunardini, \emph{{A Mathematica script for harmonic oscillator
  nuclear matrix elements arising in semileptonic electroweak interactions}},
  \href{https://doi.org/10.1016/j.cpc.2008.02.018}{\emph{Comput. Phys. Commun.}
  {\bfseries 179} (2008) 345}
  [\href{https://arxiv.org/abs/0706.2210}{{\ttfamily 0706.2210}}].

\bibitem{Bernard_2001}
V.~Bernard, L.~Elouadrhiri and U.-G. Mei{\ss}ner, \emph{Axial structure of the
  nucleon}, \href{https://doi.org/10.1088/0954-3899/28/1/201}{\emph{Journal of
  Physics G: Nuclear and Particle Physics} {\bfseries 28} (2001) R1}.

\bibitem{Donnelly:1980tsp}
T.~W. Donnelly and W.~C. Haxton, \emph{{Multipole operators in semileptonic
  weak and electromagnetic interactions with nuclei}},
  \href{https://doi.org/10.1016/0092-640X(80)90002-9}{\emph{Atom. Data Nucl.
  Data Tabl.} {\bfseries 25} (1980) 1}.

\bibitem{hnps2493}
V.~Tsakstara and T.~Kosmas, \emph{Nuclear responses to astrophysical neutrinos
  through the neutral gamow-teller strength},
  \href{https://doi.org/10.12681/hnps.2493}{\emph{HNPS Advances in Nuclear
  Physics} {\bfseries 20} (2012) 96}.

\bibitem{Gardiner:PRC2021}
S.~Gardiner, \emph{Nuclear de-excitations in low-energy charged-current
  ${\ensuremath{\nu}}_{e}$ scattering on $^{40}\mathrm{Ar}$},
  \href{https://doi.org/10.1103/PhysRevC.103.044604}{\emph{Phys. Rev. C}
  {\bfseries 103} (2021) 044604}.

\bibitem{Klos:PRD2013}
P.~Klos, J.~Men\'endez, D.~Gazit and A.~Schwenk, \emph{Large-scale nuclear
  structure calculations for spin-dependent wimp scattering with chiral
  effective field theory currents},
  \href{https://doi.org/10.1103/PhysRevD.88.083516}{\emph{Phys. Rev. D}
  {\bfseries 88} (2013) 083516}.

\bibitem{Arcadi:2019hrw}
G.~Arcadi, C.~D\"oring, C.~Hasterok and S.~Vogl, \emph{{Inelastic dark matter
  nucleus scattering}},
  \href{https://doi.org/10.1088/1475-7516/2019/12/053}{\emph{JCAP} {\bfseries
  12} (2019) 053} [\href{https://arxiv.org/abs/1906.10466}{{\ttfamily
  1906.10466}}].

\bibitem{Huh:2007zw}
J.-H. Huh, J.~E. Kim, J.-C. Park and S.~C. Park, \emph{{Galactic 511 keV line
  from MeV milli-charged dark matter}},
  \href{https://doi.org/10.1103/PhysRevD.77.123503}{\emph{Phys. Rev.}
  {\bfseries D77} (2008) 123503}
  [\href{https://arxiv.org/abs/0711.3528}{{\ttfamily 0711.3528}}].

\bibitem{Pospelov:2007mp}
M.~Pospelov, A.~Ritz and M.~B. Voloshin, \emph{{Secluded WIMP Dark Matter}},
  \href{https://doi.org/10.1016/j.physletb.2008.02.052}{\emph{Phys. Lett.}
  {\bfseries B662} (2008) 53}
  [\href{https://arxiv.org/abs/0711.4866}{{\ttfamily 0711.4866}}].

\bibitem{Chun:2010ve}
E.~J. Chun, J.-C. Park and S.~Scopel, \emph{{Dark matter and a new gauge boson
  through kinetic mixing}},
  \href{https://doi.org/10.1007/JHEP02(2011)100}{\emph{JHEP} {\bfseries 02}
  (2011) 100} [\href{https://arxiv.org/abs/1011.3300}{{\ttfamily 1011.3300}}].

\bibitem{Batell:2014yra}
B.~Batell, P.~deNiverville, D.~McKeen, M.~Pospelov and A.~Ritz,
  \emph{{Leptophobic Dark Matter at Neutrino Factories}},
  \href{https://doi.org/10.1103/PhysRevD.90.115014}{\emph{Phys. Rev. D}
  {\bfseries 90} (2014) 115014}
  [\href{https://arxiv.org/abs/1405.7049}{{\ttfamily 1405.7049}}].

\bibitem{Kaplan:2009ag}
D.~E. Kaplan, M.~A. Luty and K.~M. Zurek, \emph{{Asymmetric Dark Matter}},
  \href{https://doi.org/10.1103/PhysRevD.79.115016}{\emph{Phys. Rev. D}
  {\bfseries 79} (2009) 115016}
  [\href{https://arxiv.org/abs/0901.4117}{{\ttfamily 0901.4117}}].

\bibitem{Bi:2009uj}
X.-J. Bi, X.-G. He and Q.~Yuan, \emph{{Parameters in a class of leptophilic
  models from PAMELA, ATIC and FERMI}},
  \href{https://doi.org/10.1016/j.physletb.2009.06.009}{\emph{Phys. Lett. B}
  {\bfseries 678} (2009) 168}
  [\href{https://arxiv.org/abs/0903.0122}{{\ttfamily 0903.0122}}].

\bibitem{Kim:2015fpa}
J.-C. Park, J.~Kim and S.~C. Park, \emph{{Galactic center GeV gamma-ray excess
  from dark matter with gauged lepton numbers}},
  \href{https://doi.org/10.1016/j.physletb.2015.11.035}{\emph{Phys. Lett. B}
  {\bfseries 752} (2016) 59}
  [\href{https://arxiv.org/abs/1505.04620}{{\ttfamily 1505.04620}}].

\bibitem{Foldenauer:2018zrz}
P.~Foldenauer, \emph{{Light dark matter in a gauged $U(1)_{L_\mu-L_\tau}$
  model}}, \href{https://doi.org/10.1103/PhysRevD.99.035007}{\emph{Phys. Rev.
  D} {\bfseries 99} (2019) 035007}
  [\href{https://arxiv.org/abs/1808.03647}{{\ttfamily 1808.03647}}].

\bibitem{Dutta:2019fxn}
B.~Dutta, S.~Ghosh and J.~Kumar, \emph{{A sub-GeV dark matter model}},
  \href{https://arxiv.org/abs/1905.02692}{{\ttfamily 1905.02692}}.

\bibitem{Dutta:2020prl}
B.~Dutta, D.~Kim, S.~Liao, J.-C. Park, S.~Shin and L.~E. Strigari, \emph{Dark
  matter signals from timing spectra at neutrino experiments},
  \href{https://doi.org/10.1103/PhysRevLett.124.121802}{\emph{Phys. Rev. Lett.}
  {\bfseries 124} (2020) 121802}.

\bibitem{Johnson:2018hrx}
C.~W. Johnson, W.~E. Ormand, K.~S. McElvain and H.~Shan, \emph{{BIGSTICK: A
  flexible configuration-interaction shell-model code}},
  \href{https://arxiv.org/abs/1801.08432}{{\ttfamily 1801.08432}}.

\bibitem{Johnson:2013bna}
C.~W. Johnson, W.~E. Ormand and P.~G. Krastev, \emph{{Factorization in
  large-scale many-body calculations}},
  \href{https://doi.org/10.1016/j.cpc.2013.07.022}{\emph{Comput. Phys. Commun.}
  {\bfseries 184} (2013) 2761}
  [\href{https://arxiv.org/abs/1303.0905}{{\ttfamily 1303.0905}}].

\bibitem{ca05}
E.~Caurier, G.~Martinez-Pinedo, F.~Nowacki, A.~Poves and A.~P. Zuker, \emph{The
  shell model as a unified view of nuclear structure}, {\emph{{Reviews of
  Modern Physics}} {\bfseries 77} (2005) 427}.

\bibitem{bloom1984gamow}
S.~Bloom, \emph{Gamow-teller strength functions with the lanczos algorithm},
  {\emph{Progress in Particle and Nuclear Physics} {\bfseries 11} (1984) 505}.

\bibitem{BloomFuller1985}
S.~D. {Bloom} and G.~M. {Fuller}, \emph{{Gamow-Teller electron capture strength
  distributions in stars: Unblocked iron and nickel isotopes}},
  \href{https://doi.org/10.1016/0375-9474(85)90243-X}{\emph{Nucl. Phys. A.}
  {\bfseries 440} (1985) 511}.

\bibitem{Brown:PRC2001}
B.~A. Brown, N.~J. Stone, J.~R. Stone, I.~S. Towner and M.~Hjorth-Jensen,
  \emph{Magnetic moments of the ${2}_{1}^{+}$ states around
  $^{132}\mathrm{Sn}$},
  \href{https://doi.org/10.1103/PhysRevC.71.044317}{\emph{Phys. Rev. C}
  {\bfseries 71} (2005) 044317}.

\bibitem{Prados:PRC2007}
F.~M. Prados~Est\'evez, A.~M. Bruce, M.~J. Taylor, H.~Amro, C.~W. Beausang,
  R.~F. Casten, J.~J. Ressler, C.~J. Barton, C.~Chandler and G.~Hammond,
  \emph{Isospin purity of $t=1$ states in the $a=38$ nuclei studied via
  lifetime measurements in $^{38}\mathrm{K}$},
  \href{https://doi.org/10.1103/PhysRevC.75.014309}{\emph{Phys. Rev. C}
  {\bfseries 75} (2007) 014309}.

\bibitem{Nowacki:PRC2009}
F.~Nowacki and A.~Poves, \emph{New effective interaction for
  $0\mathit{\ensuremath{\hbar}}\ensuremath{\omega}$ shell-model calculations in
  the $\mathit{sd}\text{\ensuremath{-}}\mathit{pf}$ valence space},
  \href{https://doi.org/10.1103/PhysRevC.79.014310}{\emph{Phys. Rev. C}
  {\bfseries 79} (2009) 014310}.

\bibitem{Nummela:PRC2001}
S.~Nummela, P.~Baumann, E.~Caurier, P.~Dessagne, A.~Jokinen, A.~Knipper,
  G.~Le~Scornet, C.~Mieh\'e, F.~Nowacki, M.~Oinonen, Z.~Radivojevic,
  M.~Ramdhane, G.~Walter and J.~\"Ayst\"o, \emph{Spectroscopy of
  ${}^{34,35}\mathrm{Si}$ by $\ensuremath{\beta}$ decay: $sd\ensuremath{-}fp$
  shell gap and single-particle states},
  \href{https://doi.org/10.1103/PhysRevC.63.044316}{\emph{Phys. Rev. C}
  {\bfseries 63} (2001) 044316}.

\bibitem{coherent2018}
D.~Akimov et~al., \emph{Coherent 2018 at the spallation neutron source},  2018.
\newblock 10.48550/ARXIV.1803.09183.

\bibitem{NDS}
\emph{Live chart of nuclides. nuclear structure and decay data},  2022.
\newblock https://www-nds.iaea.org/relnsd/vcharthtml/VChartHTML.html.

\end{thebibliography}\endgroup

\end{document}